\documentclass[journal]{IEEEtran}
\ifCLASSINFOpdf
  \usepackage[pdftex]{graphicx}
\else
  \usepackage[dvips]{graphicx}
\fi
\usepackage{cite}
\usepackage{float}
\usepackage{caption}
\usepackage{hyperref}
\usepackage{booktabs}
\usepackage{comment}
\usepackage[table, x11names]{xcolor}

\begin{document}
\title{Parallel Synthesis for Autoregressive Speech Generation}
\author{Po-chun~Hsu,
        Da-rong~Liu,
        Andy T. Liu,
        and~Hung-yi~Lee
\thanks{P.-C. Hsu, D.-R. Liu, A.-T. Liu, and H.-Y. Lee are with the Graduate Institute of Communication Engineering, College of Electrical Engineering and Computer Science, National Taiwan University, Taipei 10617, Taiwan (e-mail: f07942095@ntu.edu.tw; f07942148@ntu.edu.tw; f07942089@ntu.edu.tw; hungyilee@ntu.edu.tw).}
}

\markboth{Journal of \LaTeX\ Class Files,~Vol.~14, No.~8, August~2015}
{Shell \MakeLowercase{\textit{et al.}}: Bare Demo of IEEEtran.cls for IEEE Journals}

\maketitle
\begin{abstract}
Autoregressive neural vocoders have achieved outstanding performance and are widely used in speech synthesis tasks such as text-to-speech and voice conversion. An autoregressive vocoder predicts a sample at some time step conditioned on those at previous time steps. Though it can generate highly natural human speech, the iterative generation inevitably makes the synthesis time proportional to the utterance length, leading to low efficiency. Many works were dedicated to generating the whole speech time sequence in parallel and then proposed GAN-based, flow-based, and score-based vocoders. This paper proposed a new thought for the autoregressive generation. Instead of iteratively predicting samples in a time sequence, the proposed model performs frequency-wise autoregressive generation (FAR) and bit-wise autoregressive generation (BAR) to synthesize speech. In FAR, a speech utterance is first split into different frequency subbands. The proposed model generates a subband conditioned on the previously generated one. A full-band speech can then be reconstructed from these generated subbands. Similarly, in BAR, an 8-bit quantized signal is generated iteratively from the first bit. By redesigning the autoregressive method to compute in domains other than the time domain, the number of iterations in the proposed model is no longer proportional to the utterance's length but to the number of subbands/bits. The inference efficiency is hence significantly increased. Besides, a post-filter is employed to sample audio signals from output posteriors, and its training objective is designed based on the characteristics of the proposed autoregressive methods. The experimental results show that the proposed model can synthesize speech faster than real-time without GPU acceleration. Compared with the baseline autoregressive and non-autoregressive vocoders, the proposed model achieves better MUSHRA results and shows good generalization ability while synthesizing 44 kHz speech or utterances from unseen speakers.
\end{abstract}

\IEEEpeerreviewmaketitle
\vspace{-9pt}
\section{Introduction}
\label{sec:int}
\vspace{-2pt}

\begin{figure}
  \centering
  \includegraphics[width=.93\linewidth]{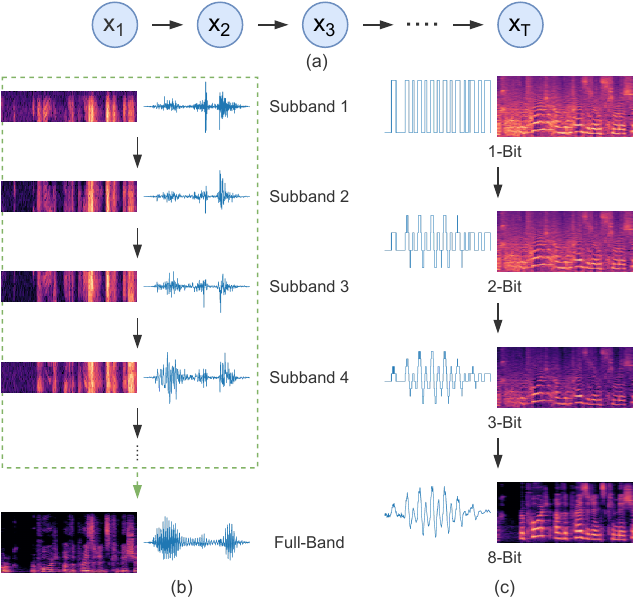}
  \vspace{-7pt}
  \caption{Overview of different autoregressive methods and their orders of generation. (a) Conventional autoregressive generation. Samples at different time steps are generated sequentially. (b) Frequency-wise autoregressive generation (FAR). Subbands are first generated autoregressively and combined to form the full-band waveform. (c) Bit-wise autoregressive generation (BAR). The spectrograms in (b) and Mel-spectrograms in (c) are only for visualization and not for generation.}
  \label{fig:overview}
  \vspace{-18pt}
\end{figure}

\IEEEPARstart{N}{eural} speech synthesis has recently achieved remarkable audio qualities in different speech tasks such as text-to-speech (TTS)~\cite{elias2021parallel, ren2020fastspeech, oord2016wavenet} and voice conversion (VC)~\cite{qian2020unsupervised}. These systems typically comprise two separate models: a synthesizer and a vocoder. A synthesizer is usually designed for some specific speech task and outputs acoustic features such as linear-scaled spectrograms, Mel-spectrograms, F0 frequencies, spectral envelopes, or aperiodicity information~\cite{wang2017tacotron, shen2018natural, ze2013statistical, tokuda2013speech}. A vocoder is designed to reconstruct audio waveforms from the acoustic features~\cite{oord2016wavenet, kalchbrenner2018efficient, morise2016world}. In the early era of neural speech synthesis, a hand-crafted vocoder~\cite{morise2016world, kawahara2001aperiodicity, wu2016merlin} or the Griffin-Lim algorithm~\cite{griffin1984signal} was adopted to reconstruct speech waveforms~\cite{wang2017tacotron, chou2018multi, yeh2018rhythm, liu2019unsupervised}. However, speech signals have a high temporal resolution, and neither conventional vocoders nor heuristic methods can reconstruct high-quality natural speech, remaining modeling raw audio a challenging problem. In Tacotron 2~\cite{shen2018natural}, a neural network~\cite{oord2016wavenet} is used as the vocoder to generate speech conditioned on Mel-spectrograms and has shown the potential of neural vocoder for synthesizing natural human speech.

Neural vocoders, while proposed for the TTS system, are not limited to this usage. A neural vocoder can be used for a wide range of applications associated with speech synthesis, such as TTS~\cite{wang2017tacotron, ren2020fastspeech, chien2021investigating}, VC~\cite{qian2020unsupervised, qian2019autovc}, speech bandwidth extension (SBE)~\cite{ling2018waveform}, and speech compression (SC)~\cite{kleijn2021generative, polyak2021speech}. With high-quality vocoders involved, these models can focus only on processing acoustic features for different speech tasks instead of directly manipulating speech waveforms. As more and more models process and output acoustic features while using another neural network to reconstruct speech, the research about using neural vocoders for waveform modeling becomes influential on the naturalness of generated speech~\cite{lorenzo2018towards, hsu2019towards, jiao2021universal}.

WaveNet~\cite{oord2016wavenet}, one of the earliest neural vocoders, adopted an autoregressive model to generate raw waveforms. The model is conditioned on previously generated samples to predict that of the next time step. It is capable of generating high-quality speech and music samples. However, due to the properties of the autoregressive method, these models are inevitably slow and inefficient while iteratively predict audio samples, restricting the speech synthesis systems from real-time applications. 
Different variants of autoregressive vocoders are proposed to increase efficiency. These works speed up the inference by leveraging more lightweight and compact architectures~\cite{jin2018fftnet, kanagawa2020lightweight, kalchbrenner2018efficient, valin2019lpcnet}, model compression techniques~\cite{kalchbrenner2018efficient, valin2019lpcnet}, domain knowledge~\cite{okamoto2017subband, okamoto2018improving, yu2019durian, valin2019lpcnet, tian2020featherwave, cui2020efficient}, specially designed generation methods~\cite{kalchbrenner2018efficient, vipperla2020bunched}, or highly optimized implementations.

Recently, significant efforts have been dedicated to the development of non-autoregressive models to better improve the inference efficiency of neural vocoders~\cite{oord2018parallel, yamamoto2020parallel, ping2018clarinet, prenger2019waveglow, kim2018flowavenet, ping2020waveflow, kumar2019melgan, kong2020hifi, chen2020wavegrad, kong2020diffwave}. Non-autoregressive models try to generate samples at different time steps in parallel. More specifically, the number of iterations in autoregressive methods is proportional to the speech length, while non-autoregressive models generate a full-length utterance in one computation. Although non-autoregressive models are faster at inference, they often underperform compared to autoregressive models in subjective evaluations~\cite{kim2018flowavenet, ping2020waveflow, hsu2020wg, kong2020diffwave, kong2020hifi, kumar2019melgan, chen2020wavegrad, song2021improved, gritsenko2020spectral}.

This work aims to enhance the efficiency of autoregressive neural vocoders while keeping the superior speech quality. To better explain the properties of autoregressive methods, we first explore and hypothesize the reason for the high quality and the low efficiency, then introduce the proposed methods motivated by the hypothesis.

Autoregressive methods have been successfully applied in various fields and achieved high-quality results. These methods predict targets word by word~\cite{vaswani2017attention, brown2020language, lewis2019bart}, frame by frame~\cite{wang2017tacotron, shen2018natural}, or pixel by pixel~\cite{van2016pixel, van2016conditional, yu2022scaling, nash2020polygen}. Similarly, in waveform generation, conventional autoregressive vocoders generate speech signals sample by sample~\cite{oord2016wavenet, kalchbrenner2018efficient}.
All these methods predict only a small part of the whole target in each generation, conditioned on the previous predictions. We hence hypothesize that, \textbf{in a generative task, dividing the target into multiple smaller parts and predicting one part at a time conditioned on the predicted parts reduce the complexity and difficulty of modeling real data}. The predicted parts can provide more detailed information for the subsequent generation process, resulting in better next predictions.
For speech synthesis, by leveraging the conditioning mechanism, autoregressive vocoders can better capture and model time dependencies in a waveform, leading to superior speech quality. The hypothesis above also explains the inefficiency of autoregressive methods in speech synthesis, primarily attributed to dividing the target in the time domain. This division significantly increases the computational cost at inference, forming a predicting process that iteratively generates tens of thousands of samples. Fig.~\ref{fig:overview} (a) illustrates the predicting process of the conventional autoregressive methods. To address the inefficiency problem while maintaining speech quality, we turn to explore alternative domains to divide a speech waveform for iterative generation.
Motivated by the success of subband analysis in speech synthesis~\cite{okamoto2017subband, okamoto2018improving, yu2019durian, tian2020featherwave, cui2020efficient, yang2021multi}, we first shift the focus from the time domain to the frequency domain. Some previous works have shown correlations between speech subbands~\cite{ming2002robust, mcauley2005subband}, and others, not restricted to speech processing, leveraged the subband correlation in their proposed methods~\cite{mcauley2005subband, okamoto2017subband, piao2007image, bhuiyan2014subband}. Combined with our hypothesis, the subband correlation revealed in the previous works inspires a thought of dividing and sequentially predicting signals in the frequency domain. Following the idea, we propose frequency-wise autoregressive generation (FAR). In FAR, a full-band speech utterance is divided into multiple frequency subbands by analysis filters. Autoregressively in the frequency domain, the model learns to predict a subband given the previous one. As shown in Fig.~\ref{fig:overview} (b), each subband contains partial information of the full-band waveform, and successive subbands are correlated, sharing similar time-dependent acoustic information provided in conventional autoregressive vocoders, such as energy, f0, formants, or voicing status. A subband signal can thus provide helpful information for the next subband prediction.
The FAR model generates the whole subband utterance in a single computation. As a result, the time needed to derive the full-band waveform is not dependent on the speech length. Instead, it is proportional to the number of subbands, which typically is considerably less than the speech length.

We further probe other possible domains for autoregressive generation. Inspired by the idea of dividing signals into different precision in \cite{kalchbrenner2018efficient}, we investigate autoregressive generation in the bit precision domain and propose bit-wise autoregressive generation (BAR). In BAR, each speech sample is quantized into an 8-bit representation using $\mu$-law companding transformation. The value of an 8-bit sample ranges from 0 to 255. The first bit, which represents whether the value of the speech sample is greater than 127, is predicted first. Then the second and the third bits are sequentially predicted, conditioned on the previously predicted bits. Finally, the first three bits are used to generate a complete 8-bit sample. Fig.~\ref{fig:overview} (c) shows the overview of BAR.
It is evident that even the 1-bit waveform shows clear f0 and formant contours. Since low-bit-coded signals contain partial acoustic information of the 8-bit waveform, they can serve as conditions to predict signals in other bit precision, implying the feasibility of BAR.
Similar to FAR, the BAR model also generates signals in the same bit precision in parallel, and the time for inference is only proportional to the number of precision we divided.

The novel FAR and BAR greatly enhance efficiency by minimizing the number of required iterations, yet they retain the iterative generative process to ensure the high quality of generated speech. 
Furthermore, FAR and BAR can be combined and applied to the same model. We show in experiments 
the effectiveness of combining FAR and BAR and the importance of the generation orders.
The contributions of this work are twofold. First, this paper is the first to explore and design new directions for autoregressive methods to improve efficiency. Instead of in the time domain, the proposed model conducts autoregressive speech generation in the frequency and bit precision domain. Second, a post-filter is applied for sampling from output posteriors and restores high-fidelity 16-bit samples instead of 8-bit ones. We combine the characteristics of the proposed autoregressive methods to design the training objective for the post-filter. 
To validate our hypothesis and assess the effectiveness of the proposed FAR, BAR, and post-filtering methods, we conducted comprehensive experiments comparing them with existing autoregressive and non-autoregressive vocoders. Objective and subjective experimental results consistently demonstrate that the proposed method, augmented with a grouping mechanism, achieves real-time synthesis and matches the quality of state-of-the-art neural vocoders. Furthermore, in addition to neural vocoders, the proposed methods have the potential to be applied to autoregressive models in various speech synthesis tasks and improve efficiency.

This paper is organized as follows. In Section~\ref{sec:rel}, we first review previous vocoding methods. In Section~\ref{sec:pro}, we described the proposed autoregressive frameworks, post-filter for posterior sampling, and vocoder architecture for the experiments. Section~\ref{sec:exp} details the experimental setup, and Section~\ref{sec:res} shows the results. The conclusion is given in Section~\ref{sec:con}.

\vspace{-5pt}
\section{Related Work}
\label{sec:rel}
\vspace{-2pt}
A vocoder aims to synthesize a speech utterance with given acoustic features. Before the era of neural vocoder, parametric vocoders such as STRAIGHT~\cite{kawahara2001aperiodicity} and WORLD~\cite{morise2016world} are usually designed according to signal processing techniques. The speech qualities of these methods are limited, and later neural vocoders are proposed and demonstrated their ability to synthesize natural speech.

\begin{figure*}
  \centering
  \includegraphics[width=.85\linewidth]{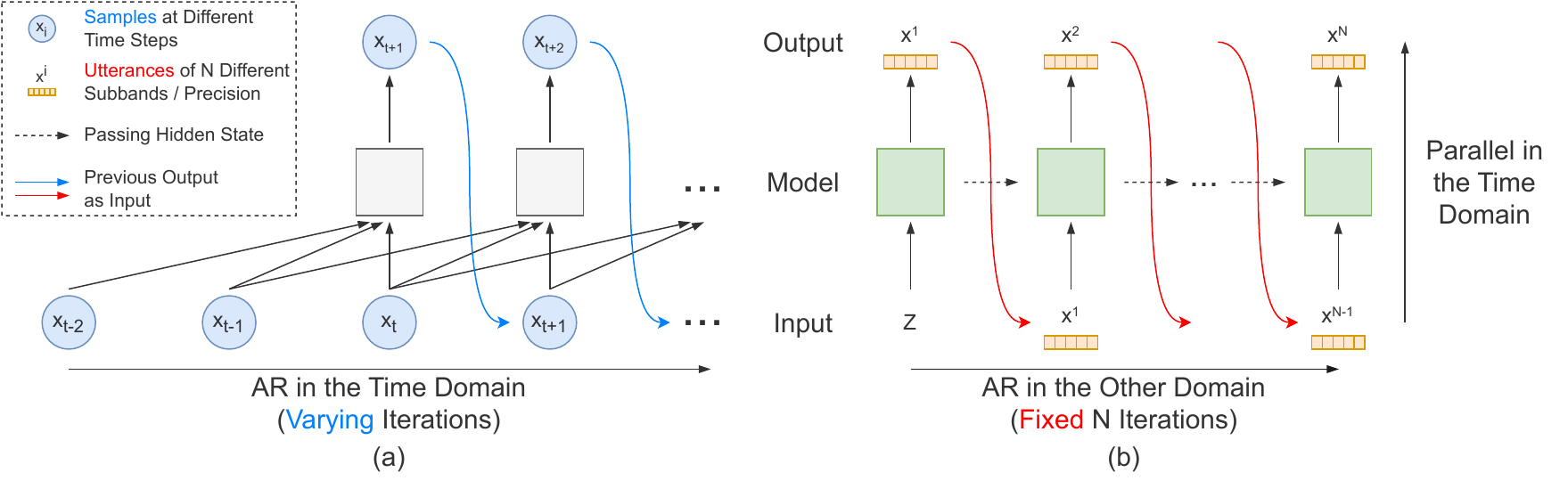}
  \vspace{-7pt}
  \caption{(a) Conventional autoregressive model. Each blue circle represents a scalar. (b) Proposed autoregressive model. The speech is generated iteratively in the frequency domain or the bit precision domain. Each green block is the model illustrated in Fig.~\ref{fig:arc} (a), and each orange block represents a time series.}
  \label{fig:ar}
  \vspace{-18pt}
\end{figure*}

\vspace{-9pt}
\subsection{Autoregressive Neural Vocoders}
\vspace{-2pt}
In the beginning, most of the neural vocoders for speech synthesis are autoregressive, meaning that they condition future audio samples on previous samples to model long-term dependencies in speech waveform. WaveNet~\cite{oord2016wavenet}, as the most representative autoregressive neural vocoder, models the waveform using dilated convolutional layers and gated activation units from \cite{oord2016conditional}. 

Some variants of autoregressive vocoders are later proposed to improve synthesis speed by leveraging more lightweight architectures and more efficient generation processes. FFTNet~\cite{jin2018fftnet} improves efficiency by an architecture resembling the Fast Fourier Transform. In WaveRNN~\cite{kalchbrenner2018efficient}, the deep convolutional networks in WaveNet are replaced with smaller recurrent neural networks to model long-term dependencies. This work also introduces Sparse WaveRNN and Subscale WaveRNN to reduce the complexity at inference time. Inheriting the compact architecture of WaveRNN, LPCNet~\cite{valin2019lpcnet} adopts linear prediction to enhance the efficiency of speech synthesis. Based on LPCNet, Bunched LPCNet~\cite{vipperla2020bunched} proposed sample bunching and bit bunching to increase inference speed, and \cite{kanagawa2020lightweight} applies tensor decomposition to reduce model parameters further.

Some works apply the subband analysis technique and significantly improve efficiency~\cite{okamoto2017subband, okamoto2018improving, yu2019durian, tian2020featherwave}. By splitting a full band waveform into several subband signals, autoregressive models can iterate on different subbands simultaneously and generate samples in parallel. Based on subband LPCNet~\cite{tian2020featherwave}, \cite{cui2020efficient} proposed to predict a subband signal conditioned on generated multiple samples from the current and other subbands.

In addition to modifying the model architecture and generation method, some other works turn from Python frameworks to highly optimized implementations in C to improve inference efficiency~\footnote{\url{https://github.com/NVIDIA/nv-wavenet}}~\footnote{\url{https://github.com/xiph/LPCNet}}. The efforts above have successfully improved the efficiency of autoregressive vocoders. However, these autoregressive models are inherently serial and can not fully utilize parallel processors like GPUs or TPUs.

\vspace{-9pt}
\subsection{Non-Autoregressive Neural Vocoders}
\vspace{-2pt}
Recently, many non-autoregressive models has been proposed to better improve the inference efficiency of the neural vocoder. Some works try to distill the output of the autoregressive model to a non-autoregressive one, for example, Parallel WaveNet~\cite{oord2018parallel} and Clarinet~\cite{ping2018clarinet}. The student networks underlying both Parallel WaveNet and Clarinet are based on inverse autoregressive flow (IAF)~\cite{kingma2016improving}. Though the IAF network can run in parallel at inference time, the teacher-student-based knowledge distillation strategy makes the training process inefficient and the whole framework complicated for users to implement.

Inspired by the success of the flow-based model in the image generation~\cite{kingma2018glow}, WaveGlow~\cite{prenger2019waveglow} is proposed. Instead of the autoregressive process in IAF, WaveGlow adopts the affine coupling layer, which is more efficient, and the synthesis speed is 25 times faster than real-time. Another group of vocoders utilizes generative adversarial networks (GAN), which calculate the probability implicitly, for example, MelGAN~\cite{kumar2019melgan} and Parallel WaveGAN~\cite{yamamoto2020parallel}. Aside from the flow-based and GAN-base models, a score-based method, such as WaveGrad~\cite{chen2020wavegrad}, models the multi-step generation as a Markov process and refines the output waveform over steps, making it possible to trade off between inference time and speech quality.

\vspace{-9pt}
\subsection{Efficiency, Quality, and Stability of Neural Vocoders}
\vspace{-2pt}
Autoregressive and non-autoregressive vocoders have different generation processes, resulting in their respective advantages and limitations regarding efficiency,  quality, and stability.

Autoregressive vocoders inherently generate samples sequentially, inevitably requiring tens of thousands of predictions to generate an utterance of a few seconds. The inference speed is thus drastically slower than real-time. Even in the faster variants\cite{jin2018fftnet, kalchbrenner2018efficient, valin2019lpcnet, vipperla2020bunched, kanagawa2020lightweight, okamoto2017subband, okamoto2018improving, yu2019durian, tian2020featherwave, cui2020efficient}, efficiency remains a concern as these models cannot generate all samples in parallel due to the nature of autoregressive generation in the time domain. In contrast, non-autoregressive models generate all signal samples in one computation, allowing for real-time synthesis even without a GPU when the architectures are compact. With the parallel synthesis capability, non-autoregressive vocoders demonstrate a significant advantage in faster inference speed compared to their autoregressive counterparts.

Even though non-autoregressive models are much faster at inference time, empirical evidence suggests that most of these methods, whether evaluated in the original papers~\cite{kim2018flowavenet, ping2020waveflow, kumar2019melgan, kong2020diffwave, chen2020wavegrad, gritsenko2020spectral} or as comparative models in other works~\cite{ping2020waveflow, hsu2020wg, kong2020diffwave, kong2020hifi, chen2020wavegrad, song2021improved, gritsenko2020spectral}, are less performant than the autoregressive baselines in subjective evaluations, indicating that non-autoregressive vocoders still struggle to consistently outperform autoregressive ones regarding speech quality. A similar tendency has been observed in natural language processing (NLP), where studies have indicated the effectiveness of autoregressive models in capturing sequential dependencies and contextual information~\cite{gu2018non, lee2020deterministic, su2021non, xiao2023survey}. Moreover, a comprehensive survey~\cite{ren2020study} on multiple tasks, including neural machine translation (NMT), automatic speech recognition (ASR), and TTS, demonstrates that the difficulty of non-autoregressive generation correlates with the target token dependency. In light of the literature above, we attribute the superior quality of autoregressive vocoders to their ability to model long-term dependencies and capture time-dependent nuances in pitch, energy, and other acoustic features. Conditioning future audio samples on previous ones, autoregressive vocoders capture temporal coherence, producing more natural speech with fine-grained acoustic details. Consequently, despite the high efficiency of non-autoregressive models, autoregressive vocoders gain an advantage in speech quality.

Besides efficiency and quality, autoregressive and non-autoregressive methods also differ in stability. While autoregressive models generally excel in speech quality, they can sometimes experience error propagation, which hampers their stability. Error propagation~\cite{wu2018beyond} refers to the phenomenon where errors in earlier predictions can propagate and amplify throughout the generation process, potentially affecting the overall stability of the output. On the other hand, non-autoregressive methods, with the ability of parallel synthesis, are not affected by this issue and offer improved stability. Though autoregressive models may encounter error propagation in some scenarios, it is worth noting that their overall quality is still better, as mentioned previously.

This paper proposes two new methods for autoregressive vocoders, FAR and BAR, to increase inference efficiency while preserving high speech quality. Compared with conventional autoregressive models, the proposed model generates signals of different subbands or bit precision iteratively and efficiently predicts samples at all time steps in parallel. The autoregressive characteristic enables the proposed model to outperform the competitive non-autoregressive models in our experiments. At the same time, the parallel generation process mitigates the impact of error propagation, leading to stable performance. Additionally, we employ a post-filter to sample audio signals from output posteriors and design a training objective based on the characteristics of the proposed methods, which has not been explored before. Besides, the proposed FAR and BAR apply not only to conventional autoregressive methods~\cite{oord2016wavenet, kalchbrenner2018efficient, valin2019lpcnet} but to non-autoregressive vocoders, such as GAN-based~\cite{yamamoto2020parallel, kong2020hifi} and flow-based~\cite{prenger2019waveglow} models. For a specific example, one can introduce FAR and BAR to Hifi-GAN to further improve speech quality, leading to similar advantages shown in \cite{morrison2022chunked} but with much higher efficiency. We leave combining the proposed efficient autoregressive methods with non-autoregressive models as a potential future work.

\vspace{-5pt}
\section{Proposed Method}
\label{sec:pro}
\vspace{-1pt}
\subsection{Rethinking the Direction for Autoregressive generation}
\label{subsec:rdaa}
\vspace{-2pt}
The conventional autoregressive neural vocoder, as shown in Fig.~\ref{fig:ar} (a), formulates the synthesis problem as maximizing the joint probability of a waveform $\textbf{x}=\{x_{1}, x_{2}, ..., x_{T}\}$, which can be factorized as follows:
\begin{equation}
  \vspace{-3pt}
  p(\textbf{x})=\prod_{t=1}^{T} p(x_{t}|x_{1}, x_{2}, ..., x_{t-1}),
  \label{eq:CAR}
  \vspace{-3pt}
\end{equation}
where $x_{t}$ is the audio sample at time $t$ and generated conditioned on signals from previous time steps. Teacher forcing can be easily applied for parallel computing during training. While at inference time, the signal $x_{t}$ at time $t$ is not able to be predicted until all $x_{1}, x_{2}, ..., x_{t-1}$ are inferred. The total iterations and time are inevitably proportional to the target audio length. Twenty-two thousand calculations are conducted iteratively to generate a one-second speech with a 22 kHz sampling rate.

To increase inference efficiency and preserve the quality of the audio output, we redesign the autoregressive method to compute in domains other than the temporal one. A waveform sequence $\textbf{x}$ is first split into $N$ subsequences, $x^{1}, x^{2}, ..., x^{N}$, where $N$ is some fixed number, and each $x^{n}$ is a time series. As shown in Fig.~\ref{fig:ar} (b), the prediction of the $n$th subsequence $x^{n}$ is conditioned on the previous subsequence $x^{n-1}$. Eq.~\ref{eq:CAR} can be reformulated as follows:
\begin{equation}
  \vspace{-3pt}
  p(\textbf{x})=\prod_{n=1}^{N} p(x^{n}|x^{n-1}).
  \label{eq:PAR}
  \vspace{-3pt}
\end{equation}
The redesigned process takes fixed $N$ iterations in total and can be parallel in the time domain by designing the model architecture, e.g., using only layers of CNN. We will discuss the splitting methods in Section~\ref{subsec:far} and \ref{subsec:bar}. Note that the conditioning acoustic features are omitted and will be detailed in Section~\ref{subsec:pva}.

\begin{figure}[t]
  \centering
  \includegraphics[width=.53\linewidth]{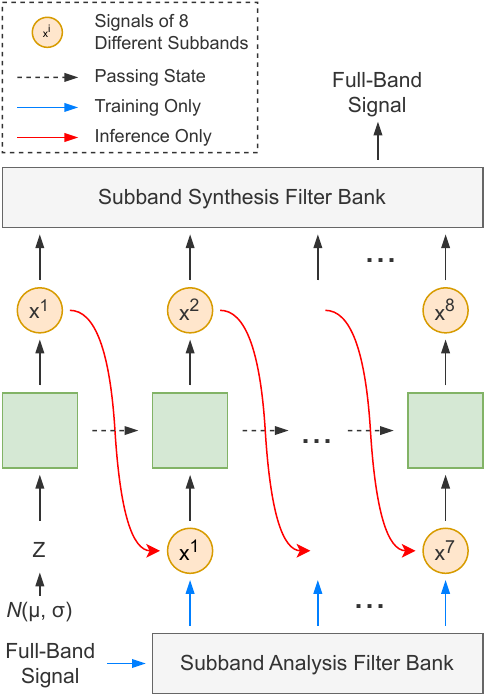}
  \vspace{-4pt}
  \caption{Frequency-wise autoregressive generation.}
  \label{fig:far}
  \vspace{-18pt}
\end{figure}

\vspace{-9pt}
\subsection{Frequency-wise Autoregressive Generation (FAR)}
\label{subsec:far}
\vspace{-2pt}
The first splitting method is subband analysis, which divides a speech utterance into multiple subband signals using a subband analysis filter bank, and each represents information in different frequency bands. The analysis process can be written as follows:
\begin{equation}
  \vspace{-3pt}
  \{x^{1}, x^{2}, ..., x^{N}\}=\phi(\textbf{x}),
  \vspace{-3pt}
\end{equation}
where $\phi$ is the analysis filter bank, $N$ is the number of subbands, and $x^{i}$ is the $i$th subband utterance. While splitting a signal with length $L$ into $N$ subband signals, each length is shortened to $L/N$. Though some works leverage this method to improve the efficiency of autoregressive vocoders~\cite{okamoto2017subband, okamoto2018improving, tian2020featherwave}, they only make use of the property of length shortening to enable models to synthesize $N$ shorter subband utterances in parallel. The inference time still grows substantially as the signal length increases.

In FAR, we replace the time domain with the frequency domain for autoregressive generation, making the model compute parallelly in the time domain and instead perform autoregressive synthesis in the frequency domain. As shown in Fig.~\ref{fig:far}, the model first takes as input a noise $Z$ from the normal distribution $N(\mu, \sigma)$ and generates the signal of the first subband, $x^{1}$. In each forward operation, $x^{i}$ serves as a condition to generate the next subband signal, $x^{i+1}$. Finally, a full-band L-length speech $\textbf{x}$ can be generated with N different subband signals and a synthesis filter bank $\psi$:
\begin{equation}
  \vspace{-3pt}
  \textbf{x}=\psi(x^{1}, x^{2}, ..., x^{N}).
  \vspace{-3pt}
\end{equation}
We follow \cite{nguyen1994near} to design analysis and synthesis filter banks. In our proposed model, a speech utterance is divided into eight subband signals, and the model iteratively generates from the subband with the highest frequency to the one with the lowest. The total number of iterations is consequently fixed to eight, which is much less than that in traditional autoregressive methods and makes the model more efficient.

\begin{figure}[t]
  \centering
  \includegraphics[width=.53\linewidth]{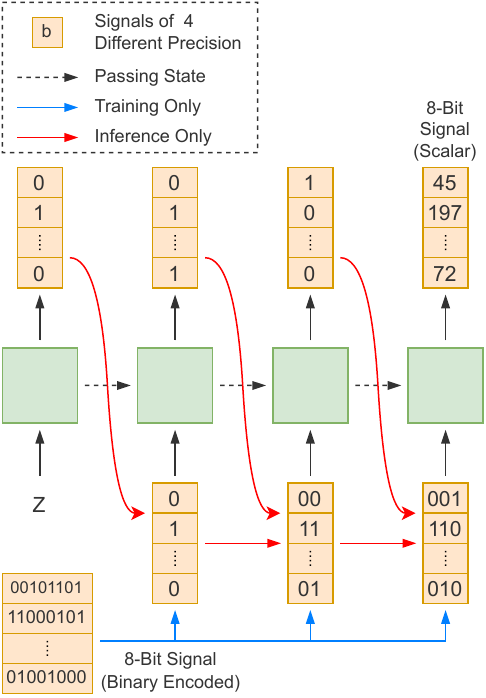}
  \vspace{-4pt}
  \caption{Bit-wise autoregressive generation.}
  \label{fig:bar}
  \vspace{-18pt}
\end{figure}

\begin{figure*}
  \centering
  \includegraphics[width=.90\linewidth]{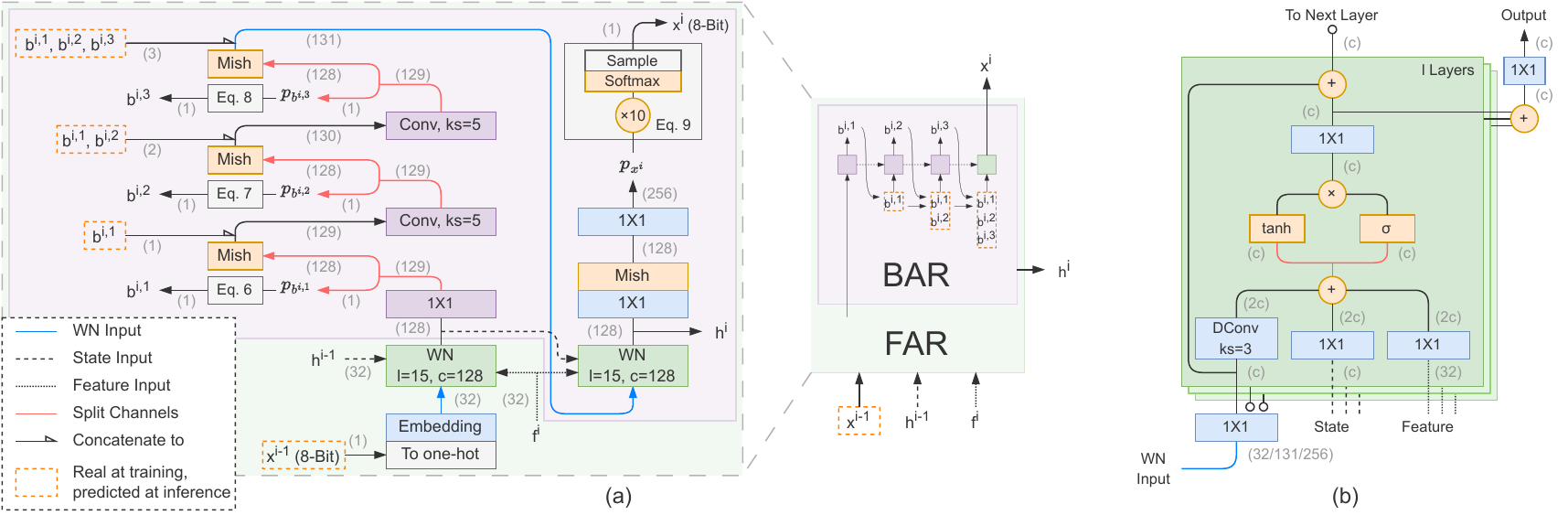}
  \vspace{-5pt}
  \caption{(a) Overview and detailed block diagram of the proposed model. The model predicts the $i$th subband $x^{i}$ conditioned on the previous subband $x^{i-1}$. $h^{i}$ is the hidden state, and $f^{i}$ is the upsampled acoustic feature. BAR is integrated in each FAR prediction, and $b^{i,1}, b^{i,2}, b^{i,3}$ are the first three bits of $x^{i}$. Channel sizes are shown in gray and parenthesized. (b) WN module.}
  \label{fig:arc}
  \vspace{-18pt}
\end{figure*}

\vspace{-9pt}
\subsection{Bit-wise Autoregressive Generation (BAR)}
\label{subsec:bar}
\vspace{-2pt}
FAR enables models to generate speech samples with a fixed number of iterations while conditioned on the previous and future information. To provide more information for autoregressive speech generation, in this section, we further investigated another domain to split the signals. The proposed BAR is an autoregressive method in the bit precision domain.

In BAR, we follow \cite{oord2016wavenet} to model the output as 8-bit samples transformed by the $\mu$-law algorithm. In a transformed 8-bit signal, samples are integers ranging from 0 to 255, denoted as 8-Bit Signal (Scalar) in Fig.~\ref{fig:bar}. Samples can also be represented as binary sequences of length eight, denoted as 8-Bit Signal (Binary Encoded). As shown in Fig.~\ref{fig:bar}, the first step is to take an initial vector $Z$ as input to generate the first bits (0 or 1) of the samples at different time steps. The first bit represents whether the value of a speech sample is greater than 127. All previous bits are used as conditions to generate the next bits. The output signal gradually becomes more precise as the bits are predicted.
Although the model can iteratively generate each bit in the binary sequences, we empirically found that the 8-bit integers in a signal can be directly predicted by conditioning solely on the first three bits, thereby fixing the total number of iterations at four. In our implementation, we integrate BAR into each FAR iteration, and the initial vector $Z$ is the hidden output in the model. Besides, we employ four separate layers for predicting the first three bits and the 8-bit integers. Further implementation details will be provided in the next section.

\vspace{-9pt}
\subsection{Proposed Vocoder Architecture}
\label{subsec:pva}
\vspace{-2pt}
In the previous sections, we introduced a new concept for autoregressive generation and demonstrated FAR and BAR. The two proposed autoregressive methods can be combined and applied to the same model. We will show in Section~\ref{subsubsec:oe-abl} that the model with both FAR and BAR outperforms the models with only either method. This section details how to combine FAR and BAR as an autoregressive neural vocoder.

Fig.~\ref{fig:arc} (a) shows an overview and a detailed block diagram of the model, consisting of two WaveNet-based modules (WN) and convolutional layers (Conv) with kernel sizes (ks) of 5 and 1 (1X1). Fig.~\ref{fig:arc} (b) shows the WN module, which is mainly composed of dilated convolutional layers (DConv) and gated activation units~\cite{oord2016conditional}. The WN module was originally proposed in WaveNet~\cite{oord2016wavenet} and has been widely adopted in many works \cite{prenger2019waveglow, yamamoto2020parallel, chorowski2019unsupervised, rethage2018wavenet}.

To conduct FAR, the model predicts the $i$th subband $x^{i}$ conditioned on the previous subband $x^{i-1}$. Each $x^{i}$ is a sequence of 8-bit integers transformed by the $\mu$-law algorithm. The input for predicting $x^{1}$ is randomly sampled from $N(0, 0.25)$ and also transformed into 8-bit integers.
The acoustic feature, a full-band Mel-spectrogram, is upsampled to the same length as the subband signals in the time domain by an upsampling network. The network consists of four convolutional layers, followed by Mish activation layers~\cite{misra2019mish} and with kernel sizes of 2, 5, 5, and 1, respectively. Nearest neighbor upsampling is applied before the second and third convolutional layers. The upsampled feature is split along the channel into eight partitions. For the $i$th subband generation, the $i$th partition is used as the conditioning feature, denoted as $f^{i}$. Inspired by \cite{kalchbrenner2018efficient}, we use $h^{i}$ as a hidden state to pass information across different iterations. All subbands are predicted by the same model with the same weights. We denote the model as $FAR$ and formulate the process as follows:
\begin{equation}
  \vspace{-3pt}
  x^{i}, h^{i}=FAR(x^{i-1}, h^{i-1}, f^{i}).
  \vspace{-3pt}
\end{equation}

BAR is integrated into each FAR iteration. In Fig.~\ref{fig:arc} (a), the first output channel of each purple blocks is used to predict the first three bits of the $i$th subband sequence $x^{i}$, denoted as $b^{i,1}$, $b^{i,2}$, and $b^{i,3}$, respectively. The remaining channels are passed to Mish activation layers and are concatenated with predicted bits (ground-truth bits at training) as input of the next layer. Conditioned on the first three bits, the 8-bit integers in $x^{i}$ are then predicted using a WN module followed by 1X1 and Mish activation layers. Note that instead of using the same WN module to predict the first three bits and the 8-bit integers, for better efficiency, we adopt three single layers to predict the first three bits, respectively. The idea of using different functions in autoregressive prediction has been introduced in \cite{xie2008dynamic, wadhvani2017review}.

Denoting the output for predicting $b^{i,1}$, $b^{i,2}$, $b^{i,3}$, and $x^{i}$ as $p_{b^{i,1}}$, $p_{b^{i,2}}$, $p_{b^{i,3}}$, and $p_{x^{i}}$, we formulate the processes of sampling from the posterior as follows:
\begin{equation}
  \vspace{-3pt}
  b^{i,1}=Sample(\sigma(10 \times p_{b^{i,1}})),
  \vspace{-2pt}
\end{equation}
\begin{equation}
  \vspace{-2pt}
  b^{i,2}=Sample(\sigma(10 \times p_{b^{i,2}})),
  \vspace{-2pt}
\end{equation}
\begin{equation}
  \vspace{-2pt}
  b^{i,3}=Sample(\sigma(5 \times p_{b^{i,3}})),
  \vspace{-2pt}
\end{equation}
\begin{equation}
  \vspace{-2pt}
  x^{i}=Sample(Softmax(10 \times p_{x^{i}})),
  \vspace{-3pt}
  \label{eq:post}
\end{equation}
where $Sample$ is randomly sampling from probabilities and can be a post-filter network when predicting $x^{i}$, which will be detailed in the next section. Similar as in \cite{okamoto2018improving}, each posterior is steepened, and the prediction becomes less noisy. We use the cross-entropy loss for training~\cite{oord2016wavenet, kalchbrenner2018efficient, valin2019lpcnet}.

\begin{figure}
  \centering
  \includegraphics[width=.63\linewidth]{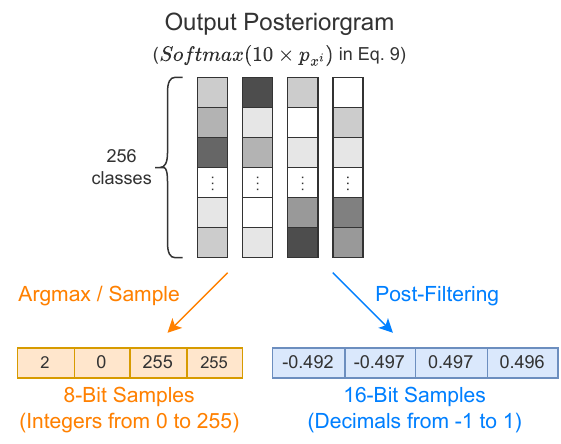}
  \vspace{-7pt}
  \caption{Different methods to sample output from posteriorgram. \textbf{Argmax}: Choosing the category with the greatest probability. \textbf{Sample}: Sampling according to the probability distribution. \textbf{Post-Filtering}: Using a network to predict 16-bit samples.}
  \label{fig:samp}
  \vspace{-18pt}
\end{figure}

\vspace{-9pt}
\subsection{Post-filtering for Posterior Sampling}
\label{subsec:pfps}
\begin{figure}
  \centering
  \includegraphics[width=.95\linewidth]{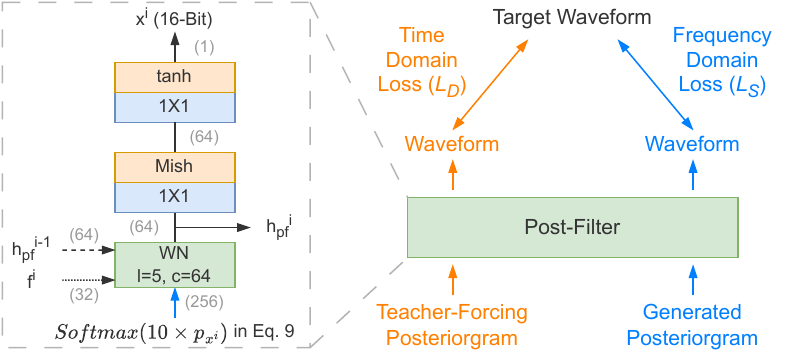}
  \vspace{-7pt}
  \caption{Post-filter and two paths for training. Channel sizes are shown in gray and parenthesized.}
  \label{fig:pft}
  \vspace{-18pt}
\end{figure}

\vspace{-2pt}
The output of a conventional autoregressive vocoder is usually a sequence of probability distributions~\cite{oord2016wavenet, kalchbrenner2018efficient, valin2019lpcnet}, and waveform signals can be generated by two different methods: (1) Selecting the category with the most significant probability (Argmax). (2) Random sampling according to the probability distributions (Sample). The orange path in Fig.~\ref{fig:samp} represents these two sampling methods. \cite{jin2018fftnet} has shown that Argmax and random sampling from the output distribution introduce noise and distortion. In \cite{jin2018fftnet} and \cite{valin2019lpcnet}, the authors leveraged voicing or pitch information and designed different rules for sampling from the distribution. While in this work, instead of carefully handcrafted sampling rules, we proposed using a neural network as a post-filter for sampling (the blue path in Fig.~\ref{fig:samp}). The proposed post-filter aims to solve two problems: (1) It reduces the effort of manually designing sampling rules by learning to restore high-quality speech from probability distributions. A well-trained post-filter can generate more accurate samples than sampling according to the distributions (2) It mitigates the noise and distortion introduced when transforming originally 16-bit samples into 8-bit representations. By directly predicting 16-bit samples, the network improves the detail and fidelity of the signal. Instead of posterior distributions of $2^{16}$ classes, the post-filter outputs the values of 16-bit samples, which are decimals ranging from -1 to 1. The process is formulated as:
\begin{equation}
  \vspace{-3pt}
  x^{i}=PF(AR(x^{i-1})),
  \vspace{-3pt}
\end{equation}
where $PF$ is the post-filter, and $x^{i}$ denotes the $i$th subband. $AR$ is the proposed autoregressive model ($h^{i-1}$ and $f^{i}$ are omitted), and the output of $AR$ is a posteriorgram, i.e., $Softmax(10 \times p_{x^{i}})$ in Eq.~\ref{eq:post}. All subbands are predicted by the same post-filter with the same weights. The architecture of the post-filter is shown in Fig.~\ref{fig:pft}.

An autoregressive model is trained under teacher-forcing manners, which means $x^{i}$ is generated conditioned on its previous ground truth subband $\hat{x}^{i-1}$. Two different types of posteriorgrams, $AR(\hat{x}^{i-1})$ and $AR(x_{ntf}^{i-1})$, can be used for training $PF$, formulated as follows:
\begin{equation}
  \vspace{-3pt}
  x_{tf}^{i}=PF(AR(\hat{x}^{i-1})),
  \vspace{-2pt}
  \label{eq:OP}
\end{equation}
\begin{equation}
  \vspace{-2pt}
  x_{ntf}^{i}=PF(AR(x_{ntf}^{i-1})),
  \vspace{-3pt}
  \label{eq:BP}
\end{equation}
where $tf$ and $ntf$ denote the sequence is generated with and without teacher forcing, respectively.
Note that at inference time, $PF$ works the same as Eq.~\ref{eq:BP}.

We train the post-filter using two different losses in the time and frequency domains, respectively. Fig.~\ref{fig:pft} shows the two training paths. In each step of the post-filter training, we perform both paths and apply different loss functions as the input type changes. When calculating the time domain loss, if the frequency domain condition (Mel-spectrogram) is the only ground truth information provided, generating a waveform close to the ground truth in the time domain is challenging. Using a teacher-forcing posteriorgram as input (the orange path) makes it more feasible to optimize the time domain loss, as partial information excluded from a Mel-spectrogram, such as phase, is provided in the ground truth input $\hat{x}^{i-1}$. The model learns to utilize this information to generate a waveform close to the real signal in the time domain. The time domain loss $L_{D}$ is written as follows:
\begin{equation}
  \vspace{-3pt}
  L_{D}=\frac{1}{N+1}(MAE(\textbf{x}_{tf}, \hat{\textbf{x}})+\sum_{i=1}^{N} MAE(x_{tf}^{i}, \hat{x}^{i})),
  \vspace{-3pt}
\end{equation}
where $\textbf{x}_{tf}=\psi(x_{tf}^{1}, x_{tf}^{2}, ..., x_{tf}^{N})$, $\hat{\textbf{x}}=\psi(\hat{x}^{1}, \hat{x}^{2}, ..., \hat{x}^{N})$, and $MAE$ is the mean absolute error.

\begin{figure}
  \centering
  \includegraphics[width=.95\linewidth]{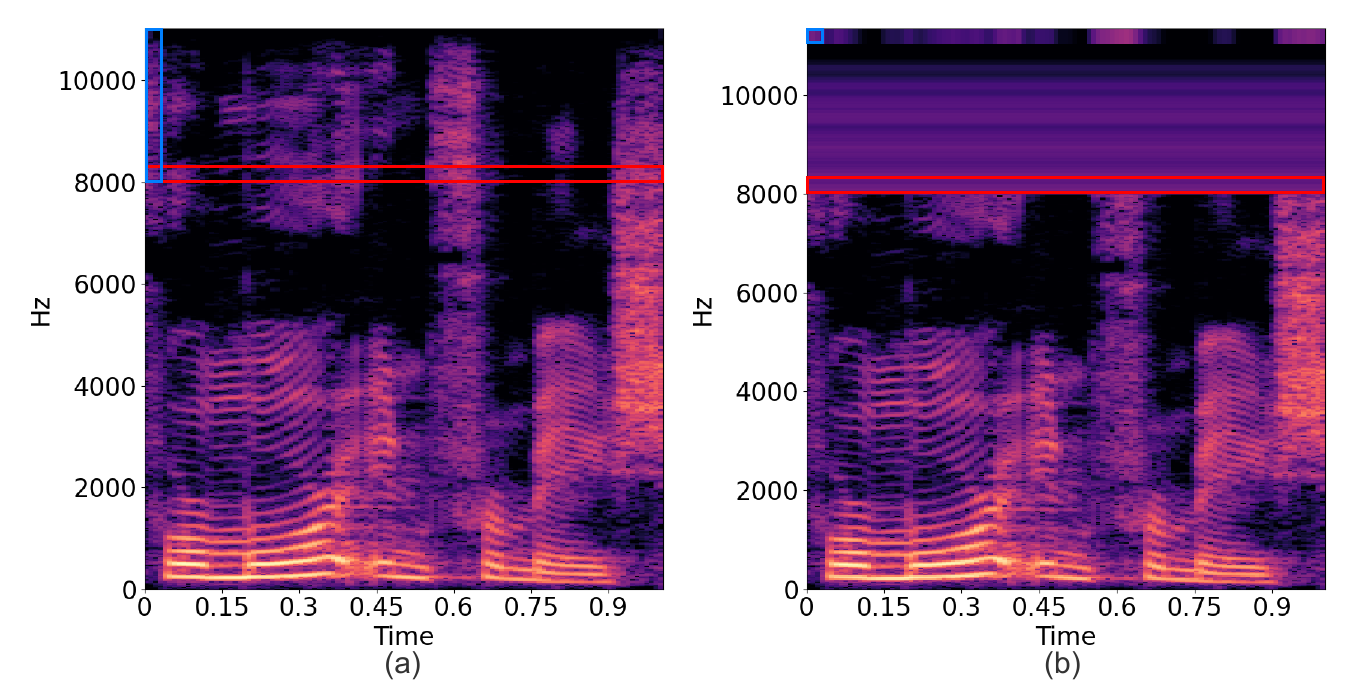}
  \vspace{-7pt}
  \caption{Average pooling in $L_{S}$. (a) STFT magnitude. (b) Band-limited STFT magnitude.}
  \label{fig:ap}
  \vspace{-18pt}
\end{figure}

For calculating the frequency domain loss, since frequency information is provided in the input Mel-spectrogram, the post-filter can take a generated posteriorgram as input (the blue path) and minimize the distance between $\textbf{x}_{ntf}$ and $\hat{\textbf{x}}$ in the frequency domain, where $\textbf{x}_{ntf}=\psi(x_{ntf}^{1}, x_{ntf}^{2}, ..., x_{ntf}^{N})$. This kind of measurement has been shown to be effective~\cite{takaki2019stft, yamamoto2020parallel, hsu2020wg, kong2020hifi}. The frequency domain loss $L_{S}$ is modified from the multi-resolution short-time Fourier transform (STFT) auxiliary loss~\cite{yamamoto2020parallel} as follows:
\begin{equation}
  \vspace{-3pt}
  L_{S}=\frac{1}{M}\sum_{m=1}^{M} (L_{sc}^{m}(\textbf{x}_{ntf}, \hat{\textbf{x}})+L_{mag}^{m}(\textbf{x}_{ntf}, \hat{\textbf{x}})),
  \vspace{-3pt}
\end{equation}
where $M$ is the number of different parameter sets of STFT; $L_{sc}$ and $L_{mag}$ are the spectral convergence loss and log STFT-magnitude loss from \cite{arik2018fast}:
\begin{equation}
  \vspace{-3pt}
  L_{sc}(a, b)=\frac{\left \|\left |STFT(a)\right |-\left |STFT(b)\right |\right \|_{F}}{\left \|\left |STFT(a)\right |\right \|_{F}},
  \vspace{-2pt}
\end{equation}
\begin{equation}
  L_{mag}(a, b)=MAE(log\left |STFT(a)\right |, log\left |STFT(b)\right |),
  \vspace{-2pt}
\end{equation}
where $\left \|\cdot\right \|_{F}$ is the Frobenius norm, and $\left |STFT(\cdot )\right |$ is the band-limited STFT magnitude. We use the magnitude information with the frequency ranging from 0 to 8000 Hz. Magnitudes with higher frequency are average pooled in the frequency and time axes to extract only local and global energy information. As shown in Fig.~\ref{fig:ap} (a), the blue box represents calculating the average per frame; the red box represents calculating the average per frequency. The frame-level averages are concatenated to each frame, and the frequency-level averages replace all values on each frequency. Fig.~\ref{fig:ap} (b) is the extracted magnitude. The blue box represents an average per frame (duplicated along the frequency axis for visualization); the red box represents an average per frequency, and the value is duplicated along the time axis for calculating $L_{S}$. The average pooling is to reduce the synthetic noise resulting from applying $L_{S}$, as mentioned in \cite{hsu2020wg, tian2020tfgan, gritsenko2020spectral}.

In each training step, the model generates both ${x}_{tf}^{i}$ and $x_{ntf}^{i}$ using the same mini-batch data. Then different loss functions, $L_{D}$ and $L_{S}$, are applied for the two cases, respectively. The final loss function $L_{PF}$ is a linear combination of $L_{D}$ and $L_{S}$:
\begin{equation}
  \vspace{-3pt}
  L_{PF}=100 L_{D}+0.1 L_{S}.
  \vspace{-3pt}
  \label{eq:LPF}
\end{equation}
We empirically decided the scalars in Eq.~\ref{eq:LPF} in preliminary experiments. A large scalar for $L_{D}$ and a small scalar for $L_{S}$ made the training more stable and led to better performance. If the scalar for $L_{S}$ is too large, the generated utterances will be with more synthetic noise mentioned in \cite{hsu2020wg, tian2020tfgan, gritsenko2020spectral}.

\vspace{-5pt}
\section{Experimental Setup}
\label{sec:exp}
\vspace{-1pt}
\subsection{Dataset}
\vspace{-2pt}
Four datasets were used in our experiments. 
\begin{itemize}
  \item \textbf{LJ Speech} LJ Speech~\cite{ljspeech17} is a high-quality speech dataset widely used for training and evaluating TTS models and neural vocoders. The audio clips are recorded with a sampling rate of 22 kHz by a female English speaker. It contains 13100 utterances, and the total duration is approximately 24 hours.

  \item \textbf{VCTK} CSTR's VCTK corpus~\cite{yamagishi2017cstr} is a multi-speaker English speech dataset containing about 44k utterances and transcriptions. Approximately 400 sentences are read by 109 English speakers with different accents. The total duration is about 44 hours. The audio clips are with a sampling rate of 48 kHz and downsampled to 22 kHz in our experiments.

  \item \textbf{CMU ARCTIC} Initially designed for unit selection speech synthesis, the CMU ARCTIC databases~\cite{kominek2004cmu} consist of clean audio clips of different English speakers. We used utterances from a male (\textit{bdl}) and a female (\textit{slt}) speaker as a test set for experiments in Section~\ref{subsec:generalization}. All the utterances were downsampled to 22 kHz.

  \item \textbf{Internal Mandarin Speech Corpus} The corpus is designed for building Mandarin TTS systems. It consists of 9004 high-quality speech utterances of a Mandarin female speaker. The total duration is about 7 hours. We used the audio clips with a sampling rate of 44 kHz for high-fidelity speech synthesis.
\end{itemize}

\vspace{-9pt}
\subsection{Acoustic Feature}
\vspace{-2pt}
We used an 80-dimensional Mel-spectrogram as the conditioning acoustic feature for speech synthesis. For computing the STFT, the FFT size, hop size and window size were set to 1024 (46 ms), 200 (9 ms), and 800 (36 ms), respectively.

\vspace{-9pt}
\subsection{Model Details}
\vspace{-2pt}
\subsubsection{Baseline and TTS Models}
Five baseline vocoder models were used in our experiments. We trained three vocoder models, including autoregressive and non-autoregressive methods. We also trained a TTS model to evaluate their performance for speech synthesis.

\begin{itemize}
  \item \textbf{WaveNet} An 8-bit WaveNet from public implementation~\footnote{\url{https://github.com/r9y9/wavenet_vocoder}} were used. The model was with 30 layers, 3 dilation cycles, 128 residual channels, 256 gate channels, and 128 skip channels. The upsampling layers for acoustic features had upsampling rates of \{5,8,5\}. Both input and output were 8-bit one-hot vectors quantized using $\mu$-law companding transformation. We trained the model with an Adam optimizer~\cite{kingma2014adam} for 500k iterations.

  \item \textbf{WaveRNN} We used the public implementation~\footnote{\url{https://github.com/fatchord/WaveRNN}} to build an 8-bit WaveRNN as a faster autoregressive baseline. The dual softmax layer and efficiency optimization techniques proposed in \cite{kalchbrenner2018efficient} were not adopted. Two GRU layers in the models had 512 channels. The upsampling layers for acoustic features had upsampling rates of \{5,8,5\}. The input and output were 8-bit quantized vectors. The network is trained with an Adam optimizer for 500k iterations.

  \item \textbf{Parallel WaveGAN} Though there were many GAN-based neural vocoders proposed recently \cite{yamamoto2020parallel, yang2020vocgan, tian2020tfgan, kong2020hifi}, we chose Parallel WaveGAN from public implementation~\footnote{\url{https://github.com/kan-bayashi/ParallelWaveGAN}} as a non-autoregressive baseline for its stable and efficient training. The upsampling rates for acoustic features were \{5,8,5\}. We followed \cite{yamamoto2020parallel} to set the parameters of the generator and discriminator. The discriminator was fixed for the first 100k steps and jointly trained with the generator for 400k steps. An Adam optimizer were used during training.

  \item \textbf{Tacotron 2} A Tacotron 2~\cite{shen2018natural} was built as a frontend for text-to-speech synthesis. The model was modified from the public implementation by NVIDIA~\footnote{\url{https://github.com/NVIDIA/tacotron2}}. We applied the reduction factor in \cite{wang2017tacotron} to improve convergence speed. The implementation is publicly available~\footnote{\url{https://github.com/BogiHsu/Tacotron2-PyTorch}}.
\end{itemize}

The other two vocoder models were LPCNet and WaveGlow. The implementations of these two models were officially released by the authors. Note that the original training settings were not exactly the same as those mentioned above. However, changing the settings without further parameter tuning may lead to suboptimal performance, and for these two models, it's not easy tuning the parameters and extensively retraining. The input of LPCNet is specially designed acoustic features~\cite{valin2019lpcnet}, including 18-dimensional BFCC and two pitch parameters (period and correlation), which are different from a Mel-spectrogram. Training a WaveGlow model takes eight GPUs and several days~\cite{prenger2019waveglow}. Hence, using the official pre-trained models is a simple and practical way to show the best performance of these methods, and we only used them in the single speaker evaluations.

\begin{table}[t]
  \centering
  \scriptsize
  \caption{Details of the multi-resolution STFT auxiliary loss. $m$ denotes the number in $L_{sc}^{m}$ and $L_{mag}^{m}$.}
  \vspace{-4pt}
  \begin{tabular}{lccc}
  \toprule
  \textbf{m}  & \textbf{1}   & \textbf{2}   & \textbf{3}  \\ \hline
  FFT size    & 2048 (93 ms) & 1024 (46 ms) & 512 (23 ms) \\
  Hop size    & 400 (18 ms)  & 200 (9 ms)   & 100 (5 ms)  \\
  Window size & 2000 (91 ms) & 1000 (45 ms) & 500 (23 ms) \\  \bottomrule
  \end{tabular}
  \label{tb:OE-STFT}
  \vspace{-18pt}
\end{table}

\begin{itemize}
  \item \textbf{LPCNet} The pre-trained LPCNet model was from the official implementation~\footnote{\url{https://github.com/xiph/LPCNet}}. The acoustic features used and the detailed parameters were set following \cite{valin2019lpcnet}. The model was implemented in Python using Keras and trained on 16 kHz utterances. At inference time, the model generated speech in either Python or C.
  
  \item \textbf{WaveGlow} The pre-trained WaveGlow model was from the official implementation~\footnote{\url{https://github.com/NVIDIA/waveglow}}, built in Python using PyTorch, and trained on LJ Speech. The model took Mel-spectrograms as input and generated 22 kHz utterances. The detailed parameters were set following \cite{prenger2019waveglow}.
\end{itemize}

\begin{table*}[ht]
  \centering
  \scriptsize
  \caption{Details and inference speed (w/o and w/ GPU) of different models. For Parallel WaveGAN, only the parameters of the generator are counted. The values in the brackets are the standard deviations.}
  \vspace{-4pt}
  \begin{tabular}{lcccccccc}
  \toprule
  \textbf{Model} & \textbf{Label} & \textbf{Type} & \textbf{Arch.} & \textbf{Loss} & \textbf{Size (M)} & \textbf{Code} & \textbf{CPU (kHz)} & \textbf{GPU (kHz)} \\ \hline
  WaveNet          & WN  & AR  & WN      & CE        & 4.7  & py & 0.19 (0.00)         & 0.13 (0.00)             \\
  WaveRNN          & WR  & AR  & RNN     & CE        & 4.3  & py & 1.14 (0.02)         & 1.87 (0.02)             \\
  LPCNet           & LPC & AR  & RNN     & CE        & 1.8  & py/c & 0.04 (0.00)/\textbf{103.7} (1.5) & 0.04 (0.00)/- \\
  WaveGlow         & WG  & NAR & WN+Flow & Flow      & 87.7 & py & 6.2 (0.2)           & \textbf{1203.8} (487.2) \\
  Parallel WaveGAN & PWG & NAR & WN      & Adv.+Aux. & 1.3  & py & 19.5 (1.5)          & \textbf{1325.8} (62.6)  \\ \hline
  Proposed         & Proposed & AR  & WN & CE+Aux.   & 5.8  & py & 8.9 (0.3)           & \textbf{393.1} (4.0)    \\
  \ -PF            & -PF  & AR  & WN     & CE        & 5.6  & py & 9.2 (0.3)           & \textbf{415.6} (3.4)    \\
  \ g-5            & g-5  & AR  & WN     & CE+Aux.   & 7.0  & py & \textbf{27.9} (0.8) & \textbf{891.6} (205.1)  \\
  \ g-10           & g-10 & AR  & WN     & CE+Aux.   & 7.3  & py & \textbf{46.3} (1.1) & \textbf{1257.0} (480.4) \\ \bottomrule
  \end{tabular}
  \label{tb:OE-COM}
  \vspace{-18pt}
\end{table*}

\begin{figure}
  \centering
  \includegraphics[width=.78\linewidth]{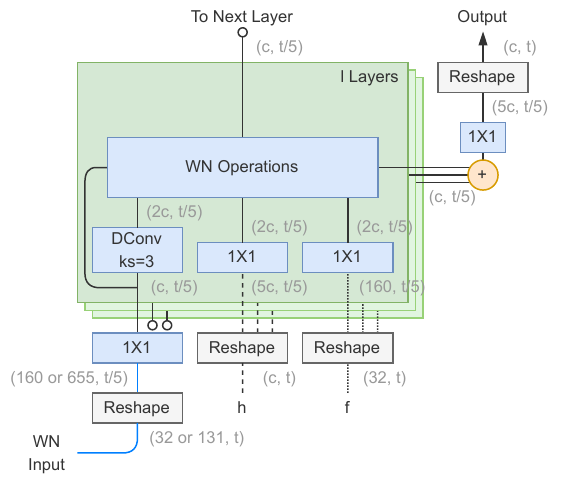}
  \vspace{-7pt}
  \caption{WN module with the grouping mechanism (g-5). Some operations are omitted for simplicity. $(c, t)$ represents a $t$-length sequence of $c$-dimensional vectors.}
  \label{fig:gm}
  \vspace{-18pt}
\end{figure}

\subsubsection{Proposed Models}
Fig.~\ref{fig:arc} and Fig.~\ref{fig:pft} show detailed parameters of the proposed model and post-filter, including kernel sizes, channel sizes, and numbers of layers. The WN modules in Fig.~\ref{fig:arc} (a) and Fig.~\ref{fig:pft} had the same dilation size growth rate. The dilation sizes of different layers grew exponentially in cycles, i.e., $[1,2,4,...,32,1,2,4,...]$. The upsampling network had upsampling rates of \{5,5\} and 80 channels, except for the output layer, which had 256 channels.

We first trained all modules but the post-filter with the cross-entropy loss for 500k steps. In each training step, the model generated subbands from the first to the eighth with ground truth previous subbands as input (i.e., the teacher-forcing mechanism). Then we fixed the weights and followed the same training process to train only the post-filter with $L_{PF}$ for 500k steps.
We set $M=3$ for calculating $L_{S}$. TABLE~\ref{tb:OE-STFT} lists the detailed parameters for the STFT. We used a batch size of 8 and the Ranger optimizer~\cite{Ranger} during training.

To further improve the synthesis speed, we applied the grouping mechanism in \cite{prenger2019waveglow} to both WN modules in Fig.~\ref{fig:arc} (a). The modified WN module is shown in Fig.~\ref{fig:gm}. The WN input sequence, hidden state, and acoustic feature are first reshaped. The channel sizes increase while the lengths are shortened inversely. We denote the model with this mechanism as g-$i$, where $i$ indicates the factor for reshaping. Note that we did not apply the grouping mechanism to the post-filter to preserve the performance.
The method was initially used to build a flow-based model for 1-D speech signals. It has shown effectiveness in improving inference efficiency since the input length is significantly reduced while the channel sizes of the hidden layers remain the same~\cite{zhai2020squeezewave, hsu2020wg}.

All the baseline, TTS, and proposed models were trained on a single NVIDIA V100 GPU using PyTorch. TABLE~\ref{tb:OE-COM} lists the details of different models. The second column shows whether the model is autoregressive (AR) or non-autoregressive (NAR). The third column describes the main architecture used to build the model, the WN module mentioned in Section~\ref{subsec:pva}, RNN models (RNN), or a specially designed flow-based model (Flow)~\cite{prenger2019waveglow}. The fourth column lists the loss functions for training, including cross-entropy loss (CE), adversarial loss (Adv.), auxiliary loss in \cite{yamamoto2020parallel} (Aux.), and specially designed loss for the flow-based model (Flow)~\cite{prenger2019waveglow}. As described in Section~\ref{subsec:pva}, we applied FAR and BAR to WaveNet to build the proposed models. Consequently, the details of the proposed models are the most similar to the WaveNet model.

\vspace{-9pt}
\subsection{Evaluation Metrics}
\vspace{-2pt}
We used four objective and one subjective metrics in the following experiments to evaluate the distortion of the generated speech in different perspectives and the quality under human perception.

\begin{itemize}
  \item \textbf{Objective Evaluation}
  We used Mel-cepstral distortion (\textbf{MCD}, reported in dB)~\cite{kubichek1993mel} to evaluate the distortion in the frequency domain. Mel-cepstrums were first extracted from ground truth and generated speech. The root mean square error (RMSE) was then calculated frame-by-frame, and the average RMSE per frame represented the distortion. To further show the pitch accuracy of generated speech, we calculated the RMSE of F0 (\textbf{F0-RMSE}, reported in Hz) and $log_{2}$F0 (\textbf{LogF0-RMSE}, reported in $10^{-2}log_{2}$Hz). We also calculated the error rate of voiced/unvoiced (V/UV) flags (\textbf{V/UV Error}, reported in \%), which was the percentage of the frames with mismatched V/UV flags. The F0 and V/UV information were extracted using pYIN algorithm~\cite{mauch2014pyin}.

  \item \textbf{Subjective Evaluation}
  We conducted modified MUltiple Stimuli with Hidden Reference and Anchor (\textbf{MUSHA}) tests~\cite{recommendation2001method} to rate the quality of generated speech under human perception. In our MUSHRA test, samples with the same speech content but generated by different systems were presented to the raters side by side. The raters then scored the samples from 0 to 100 based on their naturalness. A higher score indicates better quality in naturalness. Similar to the setup in \cite{lorenzo2018towards, jiao2021universal}, and \cite{merritt2018comprehensive, cotescu2019voice, latorre2019effect, gabrys2022voice, deja22_interspeech}, when evaluating samples with the same speech content, the raters were not forced to score at least one system with 100. We used the ground truth sample as the upper anchor and an extra system to generate the lower anchor. The system simply reconstructed the speech signal from a Mel-spectrogram using a 1-iteration Griffin-Lim algorithm~\cite{griffin1984signal}. All the raters were recruited using Amazon Mechanical Turk. The raters should self-report as native speakers (English or Mandarin, depending on the evaluation). We also asked the raters to wear headphones and stay in a quiet environment while taking the test. We randomly selected 20 utterances for the evaluation, and each utterance was scored by at least 15 subjects.
\end{itemize}

After the objective and the subjective evaluations, paired t-tests were conducted to show the statistical significance. In the ablation study, we calculated the p-values between the proposed method without $PF$ and other systems for better comparison. For the rest of the experiments, we calculated the p-values between the proposed method and other systems.

\vspace{-5pt}
\section{Results}
\label{sec:res}
\vspace{-1pt}
\subsection{Objective Evaluation}
\label{subsec:objective}
\vspace{-2pt}
In this section, we used different objective metrics to evaluate the efficiency and performance of the proposed methods and other vocoders. All the models were trained using the LJ Speech corpus. We separated 100 utterances as the test set.

\subsubsection{Model Sizes and Inference Speed}
TABLE~\ref{tb:OE-COM} shows the model sizes of different methods and their inference speed. The sixth column reports the number of model parameters (in millions). Among all methods, Parallel WaveGAN was the smallest in size. The adversarial network~\cite{goodfellow2014generative} enabled the lightweight generator in Parallel WaveGAN to learn effectively with a discriminator. WaveGlow, on the other hand, was the largest model. The flow-based architecture made the network inevitably deep~\cite{kingma2018glow, prenger2019waveglow}. The proposed model had about 30\% more parameters than WaveNet since the network was deeper. The sizes of those with the grouping mechanism (g-5 and g-10) were with more parameters since their channel sizes were 5 times and 10 times larger in the hidden layers, respectively, as shown in the blue path of Fig.~\ref{fig:gm}.

As for measuring the inference speed, we tested these models using an Intel i7-6700K CPU and an NVIDIA 2080Ti GPU. Since the efficiency of non-autoregressive models might be affected by the length of the target speech at the inference stage, we selected 8 utterances with various lengths. The duration of the audio clips ranged from 2 to 10 seconds, and the average duration was 5.8 seconds. The eighth and ninth columns of TABLE~\ref{tb:OE-COM} shows the generating rate (number of generated samples per second, in kHz) with and without GPU, respectively. We concluded our observations as follows:

\begin{itemize}
  \item
  The autoregressive models in Python generated speech slowly with and without GPU. WaveNet was a deep network with many convolutional layers. It took much computing time in each iteration and could only generate hundreds of samples per second. WaveRNN used two GRUs to replace the convolutional layers, making it faster and more lightweight, but both methods were far from real-time (22 kHz). LPCNet was built and trained in a different framework (Keras). It was inefficient while generating in Python. However, with the highly optimized implementation in C, the inference speed without GPU reached 103.7 kHz, even faster than all the other non-autoregressive methods. Since there are no other C implementations of the non-autoregressive models for comparison, the results of LPCNet are listed only to show the improvement of the C implementation over the Python implementation.
  
  \item
  The non-autoregressive methods generated utterances more efficiently with and without GPU. Parallel WaveGAN was at 1325.8 kHz and reached 60 times faster than real-time with GPU. In the case of inference only with CPU, it was still at a rate of 20 kHz. The high efficiency came from Parallel WaveGAN's lightweight generator and its parallel synthesis architecture. WaveGlow, when with GPU, reached a similar result to Parallel WaveGAN. The inference speed was much slower without GPU since the model size was larger, making the power of parallel computing more critical.
  
  \item
  The proposed model increased the autoregressive inference speed by changing from the time domain to the frequency domain and the bit precision domain. We also showed in the fourth and fifth rows that using a compact $PF$ to sample signals from posteriorgrams had only a little degradation on speed. The proposed model was 18 times faster than real-time with GPU, shortening the performance gap between autoregressive and non-autoregressive methods. Finally, we measured the rate of the proposed models with the grouping mechanism and showed the mechanism's effectiveness in improving efficiency. When inference only with CPU, g-5 and g-10 achieved 1.3 and 2.1 times faster than real-time, respectively.
  
  \item
  Unlike the autoregressive models, which had a fixed input length, the input length of the non-autoregressive models varied according to the length of the target speech. Hence, for the non-autoregressive models, the standard deviations of the inference speed were much larger, indicating that their efficiencies were affected more by the various speech lengths.
\end{itemize}

\begin{table}[t]
  \centering
  \scriptsize
  \caption{Objective evaluation results (means and standard deviations) when using ground truth acoustic features from the test set of LJ Speech. \colorbox{LemonChiffon1}{p-value $<$ 0.05}; \colorbox{Khaki1}{p-value $<$ 0.01}.}
  \vspace{-4pt}
  \begin{tabular}{lcccc}
  \toprule
  \textbf{Model} & \textbf{MCD}  & \textbf{F0-RMSE} & \textbf{LogF0-RMSE}  & \textbf{V/UV Error} \\
                 & (dB)          & (Hz)             & ($10^{-2}log_{2}$Hz) & (\%)                \\ \hline
  WN       & \cellcolor{Khaki1}3.82 (0.37) & \cellcolor{Khaki1}8.83 (16.86) & \cellcolor{Khaki1}4.51 (3.97) & \cellcolor{Khaki1}6.91 (2.45) \\
  WR       & \cellcolor{Khaki1}2.98 (0.30) & \textbf{3.97} (1.63) & \textbf{2.61} (2.00) & \cellcolor{LemonChiffon1}\textbf{4.42} (2.41) \\
  WG       & \cellcolor{Khaki1}3.16 (0.22) & \cellcolor{Khaki1}4.90 (2.38) & 3.10 (1.70) & 5.14 (2.29) \\
  PWG      & \cellcolor{Khaki1}2.41 (0.11) & \cellcolor{Khaki1}5.06 (2.67) & \cellcolor{Khaki1}3.47 (2.76) & \cellcolor{Khaki1}6.55 (3.49) \\
  Proposed & \textbf{1.93} (0.13) & 4.05 (1.66) & 2.66 (1.82) & 5.09 (3.12) \\
  \ -PF    & \cellcolor{Khaki1}2.59 (0.22) & 4.14 (1.37) & 2.67 (1.51) & 4.64 (2.18) \\
  \ g-5    & 1.94 (0.12) & 4.20 (1.09) & 2.67 (0.58) & 5.03 (2.59) \\
  \ g-10   & \cellcolor{Khaki1}2.26 (0.14) & 4.29 (1.22) & 2.73 (0.69) & 5.15 (2.58) \\ \bottomrule
  \end{tabular}
  \label{tb:OE-LJ}
  \vspace{-6pt}
\end{table}

\begin{table}[t]
  \centering
  \scriptsize
  \caption{Objective evaluation results (means and standard deviations) when using ground truth acoustic features from the test set of LJ Speech. To compare with LPCNet, ground truth and generated utterances were downsampled to 16 kHz before evaluating. \colorbox{LemonChiffon1}{p-value $<$ 0.05}; \colorbox{Khaki1}{p-value $<$ 0.01}.}
  \vspace{-4pt}
  \begin{tabular}{lcccc}
  \toprule
  \textbf{Model} & \textbf{MCD}  & \textbf{F0-RMSE} & \textbf{LogF0-RMSE}  & \textbf{V/UV Error} \\
                 & (dB)          & (Hz)             & ($10^{-2}log_{2}$Hz) & (\%)                \\ \hline
  WN       & \cellcolor{Khaki1}3.70 (0.38) & \cellcolor{LemonChiffon1}7.82 (12.70) & \cellcolor{Khaki1}4.13 (2.96) & \cellcolor{Khaki1}6.51 (3.03) \\
  WR       & \cellcolor{Khaki1}2.60 (0.29) & \cellcolor{Khaki1}\textbf{4.00} (1.92) & \cellcolor{Khaki1}\textbf{2.52} (1.16) & 4.70 (2.72) \\
  LPC      & \cellcolor{Khaki1}8.45 (0.37) & \cellcolor{Khaki1}15.87 (4.97) & \cellcolor{Khaki1}9.99 (2.00) & \cellcolor{Khaki1}17.62 (3.64) \\
  WG       & \cellcolor{Khaki1}2.11 (0.20) & 5.00 (2.12) & 3.03 (1.19) & 5.18 (2.55) \\
  PWG      & \cellcolor{Khaki1}2.45 (0.12) & 5.07 (2.21) & 3.19 (1.34) & \cellcolor{Khaki1}6.67 (2.92) \\
  Proposed & \textbf{1.91} (0.15) & 5.00 (2.88) & 3.15 (2.03) & \textbf{4.61} (2.65) \\
  \ g-5    & \textbf{1.91} (0.13) & 5.85 (6.11) & 3.34 (2.45) & 4.79 (2.61) \\
  \ g-10   & \cellcolor{Khaki1}2.24 (0.16) & 5.07 (4.91) & 2.85 (1.00) & \cellcolor{LemonChiffon1}5.23 (2.59) \\ \bottomrule
  \end{tabular}
  \label{tb:OE-LJ-16k}
  \vspace{-18pt}
\end{table}

\begin{figure*}[t]
  \centering
  \includegraphics[width=0.86\linewidth]{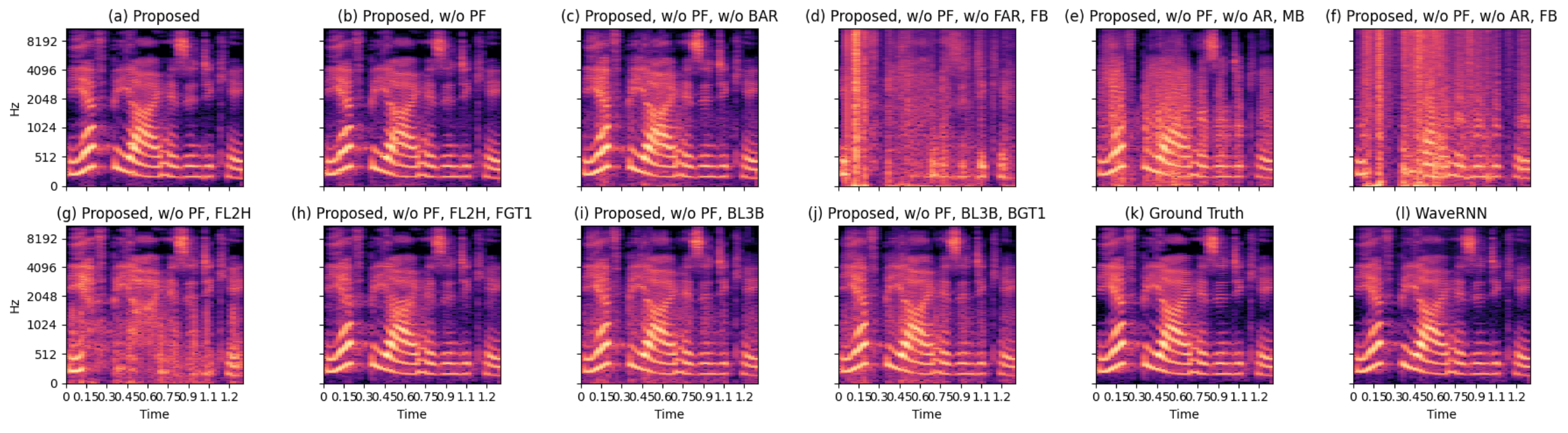}
  \vspace{-7pt}
  \caption{Mel-spectrograms of generated speech using ground truth acoustic features from LJ Speech (LJ050-0241).}
  \label{fig:ljmel}
  \vspace{-18pt}
\end{figure*}

\subsubsection{Evaluation Results}
\label{subsubsec:oe-res}
We took all the test set of LJ Speech to evaluate the models using different objective metrics. The results are listed in TABLE~\ref{tb:OE-LJ}. The proposed models (Proposed, g-5, and g-10) and Parallel WaveGAN had a lower MCD, which indicated the magnitude information was better restored. We attributed the performance to the STFT auxiliary loss~\cite{yamamoto2020parallel} since it directly optimized the magnitude distortion in the frequency domain. Regarding F0-RMSE, WaveRNN outperformed the others, but the difference was small since the values of F0 varied at a much larger scale. Besides, the difference between WaveRNN and the proposed models is not significant. Similarly, for LogF0-RMSE and V/UV Error, the results of most models were close.

Since LPCNet generated 16 kHz speech, we also downsampled the ground truth and the generated utterances to 16 kHz for another evaluation. The results are listed in TABLE~\ref{tb:OE-LJ-16k}. We found that LPCNet performed worse than the others. The distortion may be introduced when applying the block-sparse matrices to reduce the model complexity~\cite{valin2019lpcnet}.

\begin{table}[t]
  \centering
  \scriptsize
  \vspace{4pt}
  \caption{Ablation study results (means and standard deviations) when using ground truth acoustic features from the test set of LJ Speech. \colorbox{LemonChiffon1}{p-value $<$ 0.05}; \colorbox{Khaki1}{p-value $<$ 0.01}.}
  \vspace{-4pt}
  \begin{tabular}{lcccc}
  \toprule
  \textbf{Model} & \textbf{MCD}  & \textbf{F0-RMSE} & \textbf{LogF0-RMSE}  & \textbf{V/UV Error} \\
                 & (dB)          & (Hz)             & ($10^{-2}log_{2}$Hz) & (\%)                \\ \hline
  Proposed        & \cellcolor{Khaki1}\textbf{1.93} (0.13) & \textbf{4.05} (1.66) & \textbf{2.66} (1.83) & 5.09 (3.12) \\
  \ -PF           & 2.59 (0.22)  & 4.14 (1.37) & 2.67 (1.51) & \textbf{4.64} (2.18) \\ \hline\hline
  \multicolumn{5}{c}{Ablation Study on the Proposed Methods (based on -PF)} \\ \hline
  BAR-2    & \cellcolor{Khaki1}2.70 (0.21)  & 4.31 (1.80) & 2.77 (1.58) & \cellcolor{Khaki1}5.30 (2.63)          \\
  -BAR     & \cellcolor{Khaki1}4.79 (0.36)  & 4.61 (3.18) & 2.95 (2.54) & 5.00 (2.31)          \\
  -FAR, MB & \cellcolor{Khaki1}6.91 (0.58) & \cellcolor{Khaki1}7.11 (2.01) & \cellcolor{Khaki1}4.76 (1.32) & \cellcolor{Khaki1}27.28 (9.74)        \\
  -FAR, FB & \cellcolor{Khaki1}11.57 (0.54) & \cellcolor{Khaki1}9.61 (3.01) & \cellcolor{Khaki1}7.00 (1.94) & \cellcolor{Khaki1}52.54 (11.67)        \\
  -AR, MB  & \cellcolor{Khaki1}9.91 (0.54)  & \cellcolor{Khaki1}5.34 (2.86) & \cellcolor{Khaki1}3.47 (2.37) & \cellcolor{Khaki1}7.04 (3.10)          \\
  -AR, FB  & \cellcolor{Khaki1}15.41 (0.61) & \cellcolor{Khaki1}7.03 (3.17) & \cellcolor{Khaki1}4.55 (1.76) & \cellcolor{Khaki1}17.86 (7.87)         \\ \hline\hline
 \multicolumn{5}{c}{Ablation Study on the Generation Order (based on -PF)} \\ \hline
  FL2H & \cellcolor{Khaki1}6.15 (0.57) & \cellcolor{Khaki1}8.22 (6.62) & \cellcolor{Khaki1}5.50 (3.94) & \cellcolor{Khaki1}36.43 (10.75) \\
  FL2H, FGT1 & \cellcolor{Khaki1}3.68 (0.37) & \cellcolor{Khaki1}3.04 (1.13) & \cellcolor{Khaki1}1.89 (0.69) & \cellcolor{Khaki1}2.73 (1.89) \\
  FL2H, FGT2 & \cellcolor{Khaki1}1.95 (0.22) & \cellcolor{Khaki1}2.42 (1.04) & \cellcolor{Khaki1}1.48 (0.60) & \cellcolor{Khaki1}2.13 (1.85) \\
  FGT1       & \cellcolor{Khaki1}2.52 (0.22) & \cellcolor{Khaki1}3.42 (1.02) & \cellcolor{Khaki1}2.16 (0.57) & 4.66 (2.58) \\
  FGT2       & \cellcolor{Khaki1}2.54 (0.21) & \cellcolor{Khaki1}3.53 (1.44) & \cellcolor{Khaki1}2.20 (0.88) & 4.70 (2.43) \\
  FGT4mix    & \cellcolor{Khaki1}3.39 (0.29) & \cellcolor{Khaki1}3.28 (2.16) & \cellcolor{LemonChiffon1}2.12 (1.69) & \cellcolor{LemonChiffon1}4.11 (2.29) \\
  BL3B       & \cellcolor{Khaki1}6.20 (0.43) & 3.89 (1.10) & 2.46 (0.62) & \cellcolor{Khaki1}5.40 (2.52) \\
  BL3B, BGT1 & \cellcolor{Khaki1}6.28 (0.42) & 4.30 (2.16) & 2.70 (1.23) & \cellcolor{Khaki1}5.32 (2.50) \\
  BL3B, BGT2 & \cellcolor{Khaki1}5.87 (0.41) & 4.47 (4.11) & 2.80 (2.79) & 5.14 (2.53) \\
  BGT1       & \cellcolor{Khaki1}2.74 (0.38) & \cellcolor{Khaki1}2.54 (1.24) & \cellcolor{Khaki1}1.57 (0.71) & \cellcolor{Khaki1}3.74 (2.37) \\
  BGT2       & \cellcolor{Khaki1}2.49 (0.24) & \cellcolor{Khaki1}2.31 (1.29) & \cellcolor{Khaki1}1.46 (0.74) & \cellcolor{Khaki1}2.71 (2.11) \\
  -FAR, FB, BGT1       & \cellcolor{Khaki1}9.99 (0.53) & \cellcolor{Khaki1}4.63 (1.26) & 2.95 (0.72) & \cellcolor{Khaki1}8.33 (3.51) \\
  -FAR, FB, BGT2       & \cellcolor{Khaki1}6.63 (0.53) & \cellcolor{Khaki1}2.07 (0.93) & \cellcolor{Khaki1}1.28 (0.55) & \cellcolor{Khaki1}2.71 (2.04) \\ \bottomrule
  \end{tabular}
  \label{tb:OE-AS}
  \vspace{-18pt}
\end{table}

\subsubsection{Ablation Study}
\label{subsubsec:oe-abl}
We conducted a comprehensive ablation study to examine the effectiveness of the proposed methods and validate our hypothesis in Section~\ref{sec:int}. The different settings based on the proposed model without $PF$ (-PF) are denoted as follows:

\begin{itemize}
  \item \textbf{BAR-2}
  The BAR in -PF was modified to generate only the first two bits before generating the 8-bit signal.
  \item \textbf{-BAR}
  The BAR in -PF was removed. The architecture remained the intact; only each channel to predict the first three bits and to take them as input was removed.
  \item \textbf{-FAR, MB}
  The FAR in -PF was removed while keeping the architecture intact. The same model independently predicted one of the eight subbands (MB) at a time.
  \item \textbf{-FAR, FB}
  The FAR in -PF was removed while keeping the architecture intact. The model directly predicted a full-band (FB) waveform.
  \item \textbf{-AR, MB}
  The FAR and BAR in -PF were removed while keeping the architecture intact. The same model independently predicted one of the eight subbands (MB) at a time.
  \item \textbf{-AR, FB}
  The FAR and BAR in -PF were removed while keeping the architecture intact. The model directly predicted the 8-bit signal of a full-band (FB) waveform.
  \item \textbf{FL2H}
  The FAR in -PF was modified to generate subbands from the lowest to the highest frequency.
  \item \textbf{BL3B}
  The BAR in -PF was modified to generate the last three bits before generating the complete 8-bit signal.
  \item \textbf{FGT$n$ and BGT$n$}
  For $i \leq n$, after a sequence was generated in the $i$th prediction, it was replaced by the ground truth and used as input for the next prediction. For example, in BGT1, the real first bits of each subband replaced the generated ones; in FL2H, FGT1, the real lowest-frequency subband replaced the generated one to predict the next subband and to synthesize the full-band waveform.
  \item \textbf{FGT4mix}
  Similar to FGT$n$, four subbands were replaced by the ground truth for the next prediction. Instead of replacing the first four subbands, the first (the highest-frequency), third, fifth, and seventh subbands were replaced.
\end{itemize}

The results are listed in TABLE~\ref{tb:OE-AS}, and Fig.~\ref{fig:ljmel} shows the Mel-spectrograms of the generated utterances. We first discuss the effectiveness of the proposed methods and how the results support our hypothesis, concluded as follows:

\begin{itemize}
  \item
  The performance degraded when removing FAR (-PF $\rightarrow$ -FAR, MB/FB), BAR (-PF $\rightarrow$ -BAR), or $PF$ (Proposed $\rightarrow$ -PF) or when reducing the number of different bit precision in BAR (-PF $\rightarrow$ BAR-2), indicating the effectiveness of the proposed methods. Applying FAR, BAR, or PF effectively improved performance, and combining the three methods led to the best results.

  \item
  Directly predicting a waveform without autoregressive methods (-AR, FB) yielded severely distorted results, whereas dividing the target waveform into subbands (-AR, MB) improved performance. Additionally, applying an autoregressive mechanism further enhanced speech quality. Specifically, FAR (-AR, MB $\rightarrow$ -BAR) demonstrated improvements across all metrics, and BAR (-AR, MB/FB $\rightarrow$ -FAR, MB/FB) led to a better MCD. The findings above strongly support our hypothesis that dividing a target into multiple smaller parts and iteratively generating each part conditioned on predicted parts reduce prediction complexity.

  \item
  Although -FAR, MB/FB exhibited worse pitch-related results (F0-RMSE, LogF0-RMSE, and V/UV Error) than -AR, MB/FB, the degradation was due to the architecture instead of the autoregressive mechanism. In -FAR, MB/FB, 1-bit signals were first predicted without using the second WN module, as shown in Fig.~\ref{fig:arc} (a). Besides, as mentioned in Section~\ref{sec:int}, 1-bit signals contain precise F0 contours. The two facts above suggest that -FAR, MB/FB predicted much pitch information leveraging only a portion of the model capacity, leading to less accurate results. In contrast, -AR, MB/FB directly predicted the complete 8-bit signals, fully utilizing the entire model to produce pitch information. The BAR in -PF was not affected by this issue, which will be discussed together with the generation order of BAR in the subsequent paragraphs.
\end{itemize}

The ablation study results also demonstrate the importance of the generation orders in FAR and BAR.
Specifically, inverting the generation order of FAR from generating the highest-frequency subband first to the lowest-frequency one first (-PF $\rightarrow$ FL2H) resulted in increased distortion and errors. This can be explained based on the following facts:

\begin{itemize}
  \item
  Since the lower-frequency subbands contain more speech information, such as pitch, energy, and voicing status, these subbands have a greater impact on speech quality.

  \item
  In autoregressive generation, it is commonly observed that predictions in the early steps are less accurate due to the lack of information from previous predictions or hidden states. Without previous information, the first generation process is the same as predicting without an autoregressive mechanism.
\end{itemize}
Without sufficient previous information in the first generation step, the first predicted highest-frequency subband in -PF and lowest-frequency subband in FL2H were distorted. However, FL2H was much less performant since the degraded lowest-frequency subband had a more negative impact on speech quality than the degraded highest-frequency subband.

In FL2H, the lowest-frequency subband was first predicted without previous predictions, hence of low quality. The distorted lowest-frequency subband and the lack of information in the early steps would negatively affect subsequent predictions. Consequently, the distorted lower-frequency subbands led to worse speech quality. To demonstrate the importance of lower-frequency subbands, we evaluated FL2H, FGT1 and FL2H, FGT2. With ground truth low-frequency subbands provided, the improvements were significant, indicating the importance of lower-frequency subbands to high-quality speech.

Although the higher-frequency subbands in -PF were first generated and could also be distorted, the performance was less affected as these subbands contain much less speech information. Also, the improvements were not as significant as in FL2H when ground truth high-frequency subbands were provided (-PF $\rightarrow$ FGT1 and -PF $\rightarrow$ FGT2). These findings indicate that the higher-frequency subbands have minimal impact on speech quality and can be generated in the early steps.

Regarding BAR, we conducted similar experiments. Inverting the generation order from generating the first three bits first to the last three bits first (-PF → BL3B) also showed worse results, while providing the ground truth last few bits (BL3B, BGT1, and BL3B, BGT2) did not improve the performance. On the other hand, providing the ground truth first few bits (BGT1, and BGT2) led to better pitch-related results, indicating that the first few bits are more crucial for accurately predicting pitch.
The importance of the first few bits, combined with the previously mentioned lack of information issue in the early steps, provides another explanation for why -FAR, MB/FB performed worse on pitch-related metrics, which is that BAR in -FAR, MB/FB predicted low-bit-coded signals with less accurate pitch information in the early steps.
In contrast, the BAR in -PF was less affected by the lack of information. Despite generating the first three bits in the early steps, -PF still produced high-quality speech since it combined FAR and BAR. In -PF, the lower-frequency subbands, which were more critical to high-quality speech, were predicted with more acoustic information from previous subbands and hidden states. The hidden states possessed rich information about previous predictions since they were iteratively computed throughout the generation process. Sufficient information helped the BAR in -PF to predict high-quality low-bit-coded signals of the lower-frequency subbands, leading to better results.

We also observed that in BAR, the MCD increased when the real first target was given (-PF $\rightarrow$ BGT1 and BL3B $\rightarrow$ BL3B, BGT1) and decreased when the real first two targets were given (-PF $\rightarrow$ BGT2 and BL3B $\rightarrow$ BL3B, BGT2)~\footnote{The first two targets of -PF were the first two bits of the 8-bit signal, and the first two targets of BL3B were the last two bits.}. We inferred this was due to the mismatch between the consecutive subbands when the given ground truth information was insufficient. BGT1 generated a subband with only partial ground truth information (the first bits). The generated subband might still have differences, such as phase discrepancy, from the ground truth, leading to a mismatch with the ground truth first bits given in the next subband prediction. In BGT2, with more ground truth information (the first and the second bits), the generated subband was closer to the ground truth, decreasing the mismatch. The explanation above also applies to BL3B, BGT1 and BL3B, BGT2.

To verify that a mismatch may occur when consecutive subbands are partial ground truth and partial generated, we evaluated FGT4mix. In FGT4mix, a generated subband mismatched the ground truth next subband since the former was not predicted conditioned on the latter. The MCD results hence worsened despite more information being provided, indicating that the mismatch between subbands affects speech quality. Note that FGT1 and FGT2 did not suffer any mismatches since the generated parts were predicted conditioned on the ground truth subbands. Furthermore, for -FAR, FB, which applies only BAR and directly predicts a full-band waveform, there was no mismatch since the subband mechanism was removed. All metrics were improved when part of the ground truth were given (-FAR, FB $\rightarrow$ -FAR, FB, BGT1 and -FAR, FB $\rightarrow$ -FAR, FB, BGT2). These observations well support the claim that mismatches between subbands degraded speech quality.

The comprehensive ablation study shows the importance of the generation orders and explains how we decided the orders for FAR and BAR according to the results.

\begin{figure}[t]
  \centering
  \includegraphics[width=.80\linewidth]{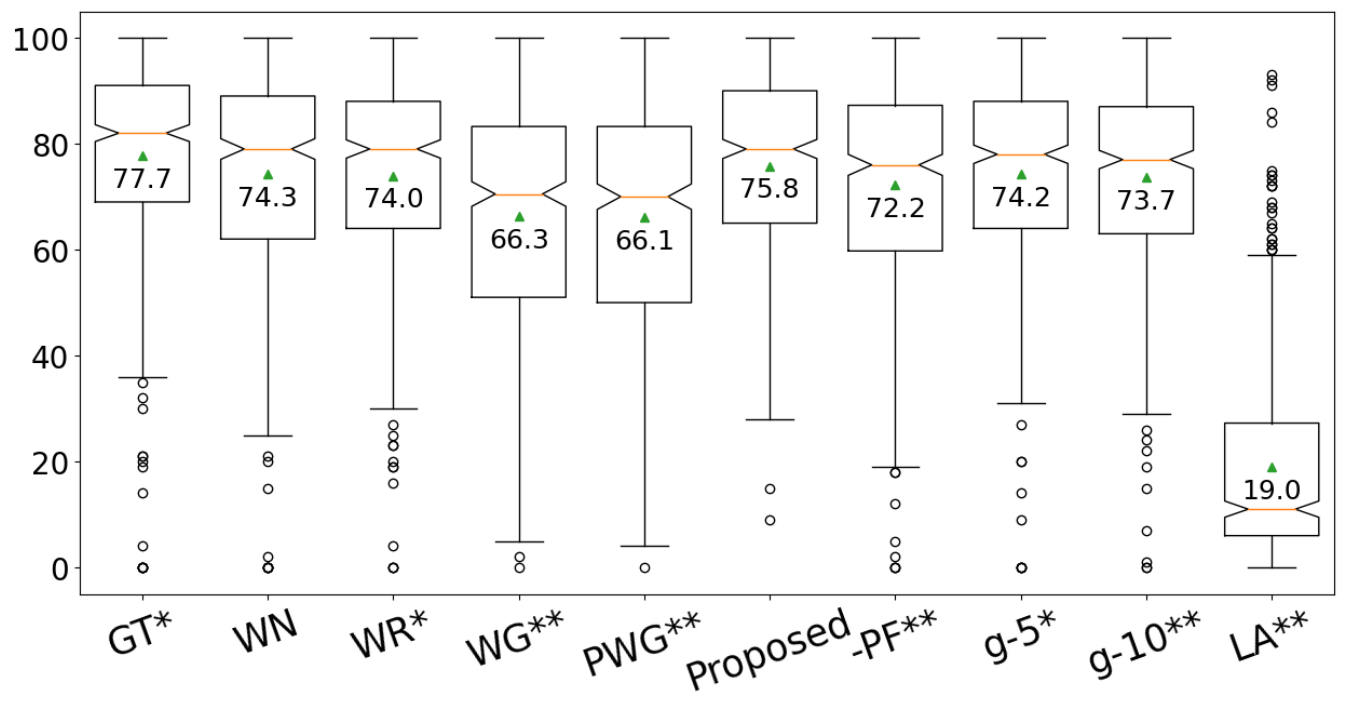}
  \vspace{-6pt}
  \caption{MUSHRA results when using ground truth acoustic features from the test set of LJ Speech. *: p-value $<$ 0.05; **: p-value $<$ 0.01.}
  \label{fig:SE-LJ}
  \vspace{-18pt}
\end{figure}

\vspace{-9pt}
\subsection{Subjective Evaluation}
\label{subsec:subjective}
\vspace{-2pt}
\subsubsection{Evaluation on Ground Truth Acoustic Features}
We randomly selected 20 utterances from the test set of LJ Speech for the subjective evaluation. The MUSHRA results of different models are listed in Fig.~\ref{fig:SE-LJ}. We concluded our observations as follows:

\begin{itemize}
  \item
  Among the baseline models, the autoregressive models bettered the non-autoregressive ones with significant differences (p-value $<$ 0.01). Besides, the baseline models with the same type (autoregressive or non-autoregressive) performed similarly, and there was no significant difference (p-value $>$ 0.05).
  
  \item
  The proposed model outperformed the others and reached a MUSHRA score of 75.8. There was no significant difference between the proposed method and WaveNet, showing that both models generated utterances closest to natural human speech, yet the proposed method reached thousands of times more efficient than WaveNet according to TABLE~\ref{tb:OE-COM}.
  
  \item
  WaveGlow and Parallel WaveGAN achieved intermediate results in the objective evaluation. However, they were still relatively low-quality to the competitive systems, and subjects in the MUSHRA test tended to give lower scores. The audio samples on the demo~\footnote{\url{https://bogihsu.github.io/TASLP2021-Parallel/Demo/demo.html}} page better show the quality of different vocoders.
  
  \item
  The score of the proposed model without $PF$ dropped by 3.6, indicating the importance of $PF$ in synthesizing natural speech. As for the grouping mechanism, the scores of g-5 and g-10 degraded slightly but still close to WaveNet and WaveRNN, showing the ability to improve the synthesis efficiency while preserving the quality.
\end{itemize}

\begin{figure}[t]
  \centering
  \includegraphics[width=.80\linewidth]{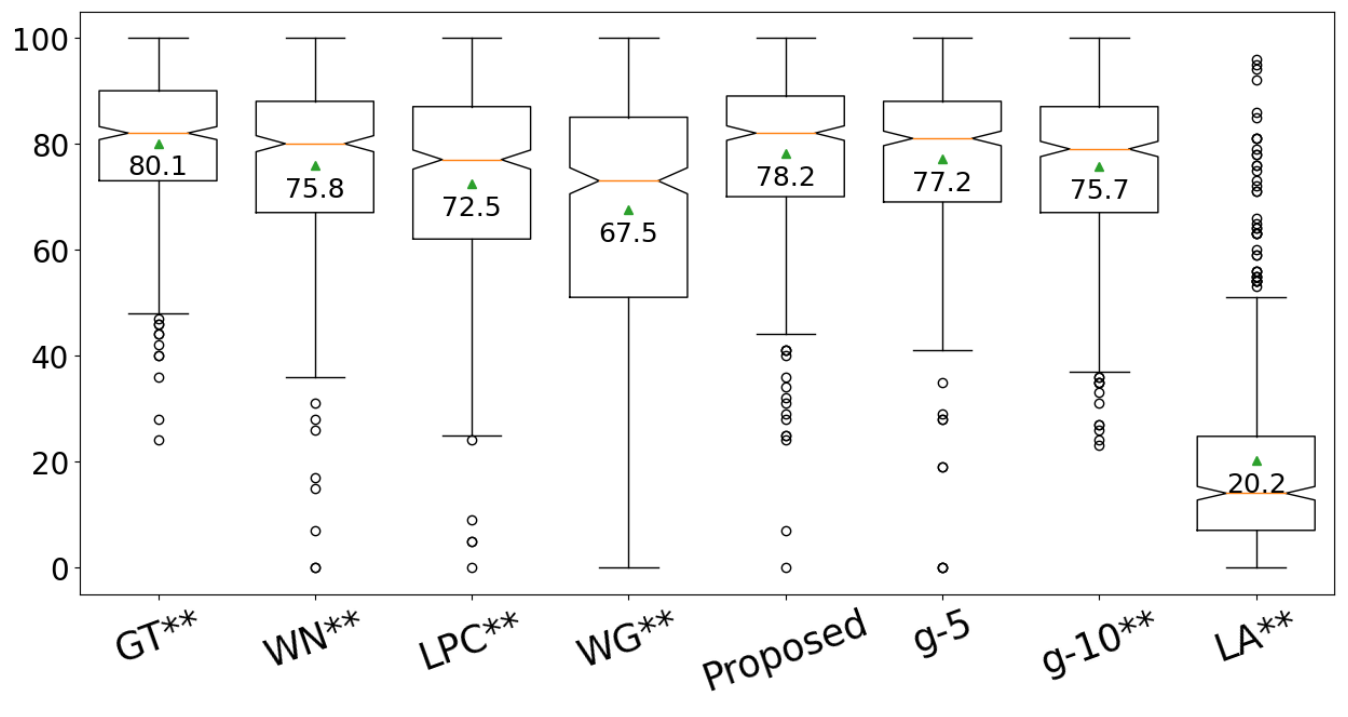}
  \vspace{-6pt}
  \caption{MUSHRA results when using ground truth acoustic features from the test set of LJ Speech. To compare with LPCNet, ground truth and generated utterances were downsampled to 16 kHz before evaluating. **: p-value $<$ 0.01.}
  \label{fig:SE-LJ-16k}
  \vspace{-6pt}
\end{figure}

\begin{figure}[t]
  \centering
  \includegraphics[width=.80\linewidth]{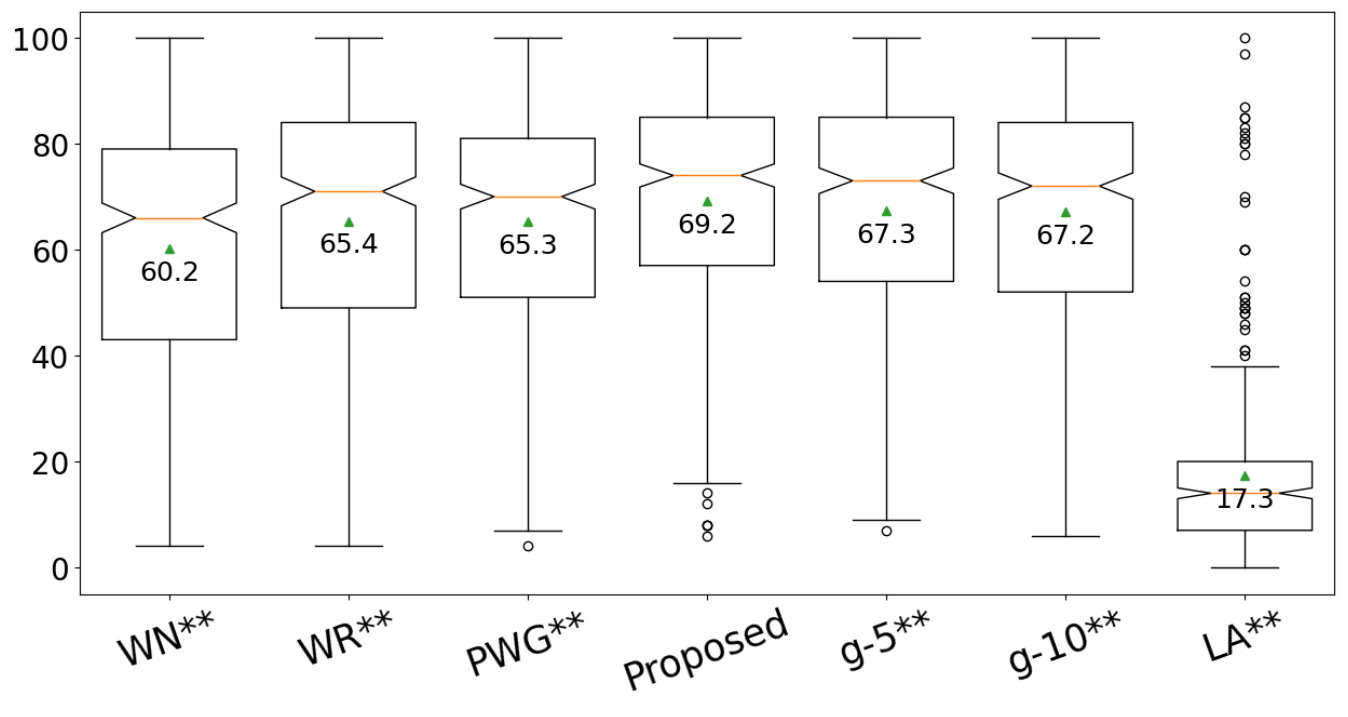}
  \vspace{-6pt}
  \caption{MUSHRA results when using acoustic features generated from Tacotron 2. **: p-value $<$ 0.01.}
  \label{fig:SE-TTS}
  \vspace{-18pt}
\end{figure}

Similar to in Section~\ref{subsubsec:oe-res}, to compare with LPCNet, we also downsampled the utterances to 16 kHz and ran another MUSHRA test. The results are listed in Fig.~\ref{fig:SE-LJ-16k}. LPCNet performed worse than WaveNet and the proposed methods but still bettered non-autoregressive  WaveGlow.

\subsubsection{Evaluation on Acoustic Features from Tacotron 2}
We combined vocoders with a Tacotron 2 model as complete TTS systems for further evaluation. The Tacotron 2 was trained using the training set of LJ Speech. During inference, we selected 20 sentences from Harvard Sentences~\cite{rothauser1969ieee} as the text input for the systems.

The results are shown in Fig.~\ref{fig:SE-TTS}. Compared with the evaluation using ground truth acoustic features, the scores between the baseline methods become closer. We found that features from Tacotron 2 had more distortion than ground truth features. In this case, WaveNet and WaveRNN were unstable and less performant. The reduced robustness of the autoregressive models to artificial acoustic features could be attributed to the error propagation from autoregressive properties in the time domain. When a suboptimal sample is generated with distorted acoustic features, it is used as the condition for the next prediction. Even if the acoustic features are clean in subsequent time steps, a flawed sample from the previous step can lead to further erroneous predictions, causing the errors to propagate and resulting in speech with noise and artifacts. 

In contrast, the non-autoregressive Parallel WaveGAN was much more stable regardless of the quality of the input features. The proposed models, possessing the properties of autoregressive and non-autoregressive methods, were not affected by error propagation, either. All three models maintained high quality and outperformed the baseline models with significant differences (p-value $<$ 0.05).

\vspace{-9pt}
\subsection{Generalization Evaluation}
\label{subsec:generalization}
\vspace{-2pt}
In this section, we studied the generalization abilities of the baseline and proposed models under different situations, including synthesizing speech of unseen speakers and high-fidelity speech with a 44 kHz sampling rate.

\subsubsection{Generalization to Unseen Speakers}
To build a multi-speakers version of the vocoders, we used the VCTK dataset for training. We use utterances from speaker \textit{bdl} and \textit{slt} in CMU ARCTIC as a male and a female test set, respectively. For each test set, we took 30 utterances (arctic\_b0500$\sim$arctic\_b0529) for objective evaluation and randomly selected 10 audio clips for subjective evaluation.

\begin{figure}[t]
  \centering
  \includegraphics[width=.80\linewidth]{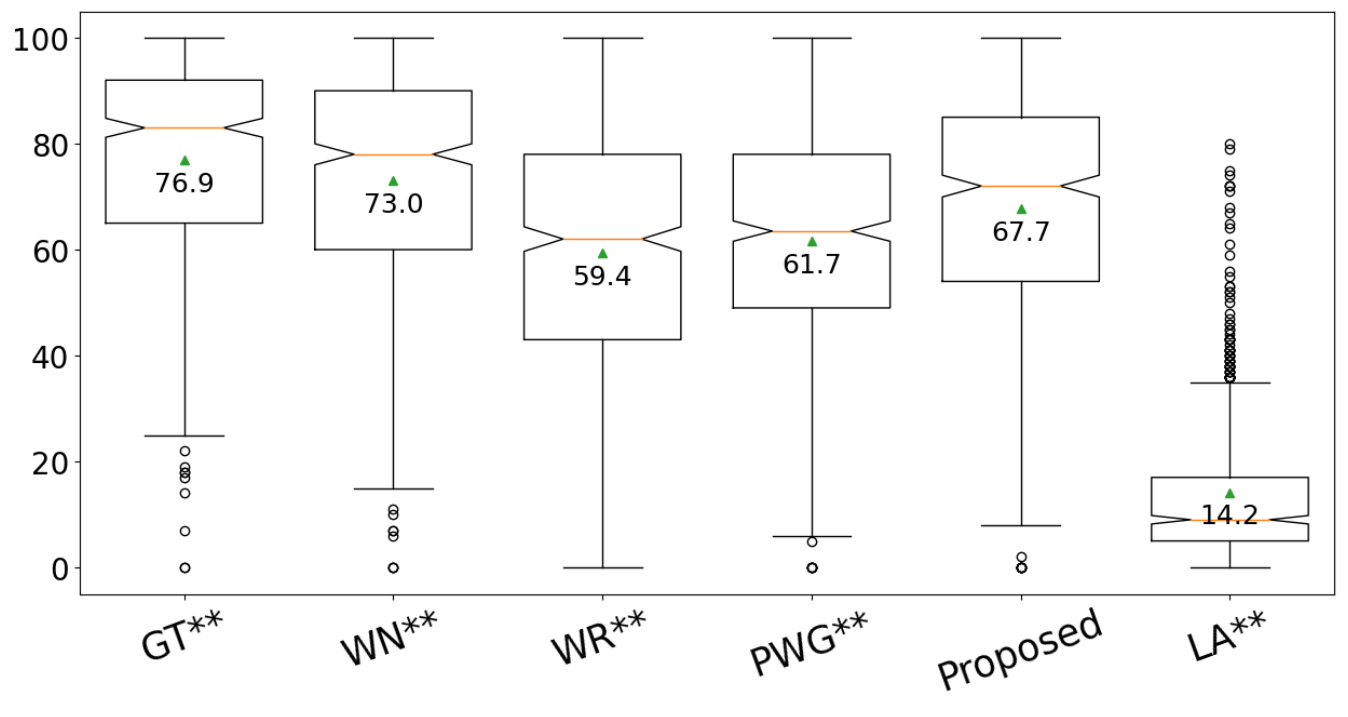}
  \vspace{-6pt}
  \caption{MUSHRA results when using ground truth acoustic features from speaker \textit{bdl} and \textit{slt} in CMU ARCTIC. The vocoders were trained on VCTK corpus. **: p-value $<$ 0.01.}
  \label{fig:SE-VCTK}
  \vspace{-18pt}
\end{figure}

TABLE~\ref{tb:OE-VCTK} and Fig.~\ref{fig:SE-VCTK} list the results. The tendency of the objective scores between different models is similar to the results in Section~\ref{subsec:objective}. WaveRNN performed slightly worse in MCD, reflected in generated speech with more noise and a lower perceptual score. The MUSHRA results showed that WaveNet generalized best. The proposed model also had a higher score than WaveRNN and Parallel WaveGAN, indicating its ability to generalize to unseen speakers.

\begin{table}[t]
  \centering
  \scriptsize
  \caption{Objective evaluation results (means and standard deviations) when using ground truth acoustic features from speaker \textit{bdl} and \textit{slt} in CMU ARCTIC. The vocoders were trained on VCTK corpus. \colorbox{LemonChiffon1}{p-value $<$ 0.05}; \colorbox{Khaki1}{p-value $<$ 0.01}.}
  \vspace{-4pt}
  \begin{tabular}{lcccc}
  \toprule
  \textbf{Model} & \textbf{MCD}  & \textbf{F0-RMSE} & \textbf{LogF0-RMSE}  & \textbf{V/UV Error} \\
                 & (dB)          & (Hz)             & ($10^{-2}log_{2}$Hz) & (\%)                \\ \hline
  \multicolumn{5}{c}{\textit{bdl} (male)} \\ \hline
  WN       & \cellcolor{Khaki1}2.58 (0.21) & 3.20 (2.52)          & 4.01 (3.61) & 8.22 (4.93) \\
  WR       & \cellcolor{Khaki1}2.64 (0.21) & \textbf{2.95} (0.66) & \textbf{3.64} (0.85) & \cellcolor{Khaki1}11.00 (6.02)         \\
  PWG      & \cellcolor{Khaki1}2.90 (0.31) & 3.68 (2.53)          & 4.62 (3.09) & 7.93 (4.62)         \\
  Proposed & \textbf{1.98} (0.23) & 3.01 (1.33) & 3.80 (1.83) & \textbf{7.22} (5.01) \\ \hline\hline
  \multicolumn{5}{c}{\textit{slt} (female)} \\ \hline
  WN       & \cellcolor{Khaki1}2.49 (0.23) & \cellcolor{LemonChiffon1}3.81 (1.37) & 2.88 (1.00) & \cellcolor{Khaki1}4.81 (3.10) \\
  WR       & \cellcolor{Khaki1}3.56 (0.33) & 3.34 (1.11) & 2.69 (0.85) & 2.79 (2.51) \\
  PWG      & \cellcolor{Khaki1}2.71 (0.15) & \cellcolor{Khaki1}4.35 (1.46) & \cellcolor{Khaki1}3.49 (1.09) & \cellcolor{LemonChiffon1}4.03 (3.19) \\
  Proposed & \textbf{2.16} (0.18) & \textbf{3.04} (1.42) & \textbf{2.42} (1.09) & \textbf{2.75} (2.39) \\ \bottomrule
  \end{tabular}
  \label{tb:OE-VCTK}
  \vspace{-6pt}
\end{table}

\subsubsection{Generalization to High-Fidelity Dataset}
To study the performance when generalization to high-fidelity dataset, we trained the vocoders on our 44 kHz internal Mandarin speech corpus.
All STFT parameters in samples remained the same as described in Section~\ref{sec:exp} and TABLE~\ref{tb:OE-STFT}, i.e., STFT parameters in milliseconds were halved in this 44 kHz experiment.
We took 30 utterances from the test set for objective evaluation and randomly selected 20 utterances for subjective evaluation.

\begin{figure}[t]
  \centering
  \includegraphics[width=.80\linewidth]{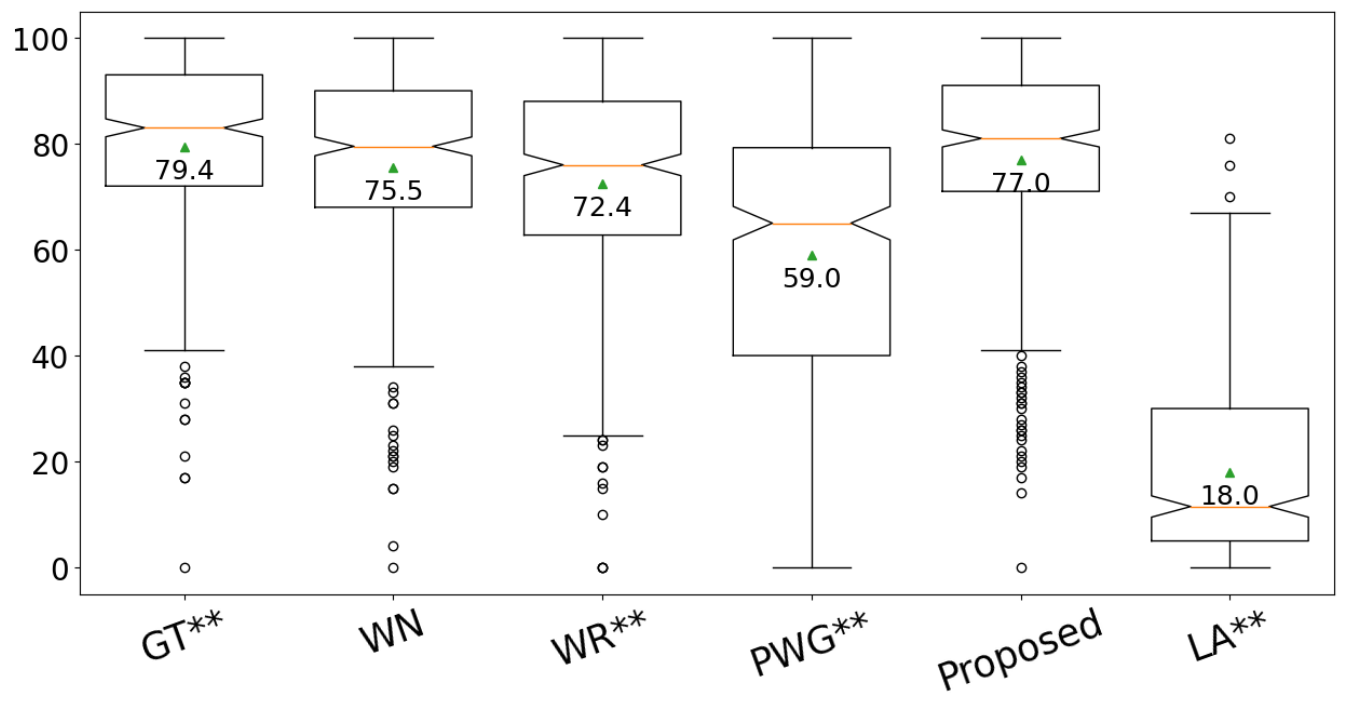}
  \vspace{-6pt}
  \caption{MUSHRA results when using ground truth acoustic features from the test set of our 44 kHz internal mandarin speech corpus. **: p-value $<$ 0.01.}
  \label{fig:SE-YT}
  \vspace{-18pt}
\end{figure}

The results are shown in TABLE~\ref{tb:OE-YT} and Fig.~\ref{fig:SE-YT}. The objective evaluation results showed a similar tendency as previous. We found that speech volume from WaveNet varied in the same utterance, leading to worse objective evaluation results. As for the MUSHRA results, utterances from WaveRNN had a bit more noise but were still graded with good scores. Besides, Parallel WaveGAN failed to generate natural results. On the other hand, both WaveNet and the proposed model achieved close scores to the ground truth, and there was no significant difference between these two systems, indicating their ability to synthesize natural high-fidelity speech.

\begin{table}[t]
  \centering
  \scriptsize
  \caption{Objective evaluation results (means and standard deviations) when using ground truth acoustic features from the test set of our 44 kHz internal mandarin speech corpus. \colorbox{LemonChiffon1}{p-value $<$ 0.05}; \colorbox{Khaki1}{p-value $<$ 0.01}.}
  \vspace{-4pt}
  \begin{tabular}{lcccc}
  \toprule
  \textbf{Model} & \textbf{MCD}  & \textbf{F0-RMSE} & \textbf{LogF0-RMSE}  & \textbf{V/UV Error} \\
                 & (dB)          & (Hz)             & ($10^{-2}log_{2}$Hz) & (\%)                \\ \hline
  WN  & \cellcolor{Khaki1}4.32 (0.88) & \cellcolor{Khaki1}31.10 (45.57) & \cellcolor{Khaki1}18.94 (19.43)          & \cellcolor{Khaki1}15.43 (6.63) \\
  WR  & \cellcolor{Khaki1}5.11 (0.37) & \textbf{4.59} (3.00) & 3.23 (3.70) & 7.12 (4.08) \\
  PWG & \cellcolor{Khaki1}3.25 (0.20) & \cellcolor{LemonChiffon1}9.32 (10.21) & \cellcolor{Khaki1}5.71 (4.48) & \cellcolor{Khaki1}11.06 (6.08) \\
  Proposed & \textbf{2.10} (0.26) & 6.70 (13.37) & \textbf{2.49} (1.10) & \textbf{6.07} (3.48) \\ \bottomrule
  \end{tabular}
  \label{tb:OE-YT}
  \vspace{-6pt}
\end{table}
  
\vspace{-9pt}
\section{Conclusion}
\label{sec:con}
\vspace{-2pt}
In this paper, we propose novel frequency-wise autoregressive generation (FAR) and bit-wise autoregressive generation (BAR). The former iteratively generates utterances of different frequency subbands, and the latter iteratively generates utterances with different bit precision. We also applied a post-filter for posterior sampling to further improve the quality. Compared with conventional autoregressive vocoders, the proposed model computes a fixed number of iterations, and the inference time is no longer proportional to the speech length. Consequently, the model possesses high inference efficiency and achieves faster than real-time without GPU acceleration. Besides, the perceptual evaluations also show that the proposed model is able to generate high-quality natural speech under different scenarios.
\vspace{-9pt}
\vspace{-9pt}
\section{Acknowledgment}
\vspace{-2pt}
We thank National Center for High-performance Computing (NCHC) of National Applied Research Laboratories (NARLabs) in Taiwan for providing computational and storage resources.
\vspace{-9pt}
\bibliographystyle{IEEEtran}
\bibliography{mybib.bib}

\begin{thebibliography}{10}
\providecommand{\url}[1]{#1}
\csname url@samestyle\endcsname
\providecommand{\newblock}{\relax}
\providecommand{\bibinfo}[2]{#2}
\providecommand{\BIBentrySTDinterwordspacing}{\spaceskip=0pt\relax}
\providecommand{\BIBentryALTinterwordstretchfactor}{4}
\providecommand{\BIBentryALTinterwordspacing}{\spaceskip=\fontdimen2\font plus
\BIBentryALTinterwordstretchfactor\fontdimen3\font minus
  \fontdimen4\font\relax}
\providecommand{\BIBforeignlanguage}[2]{{%
\expandafter\ifx\csname l@#1\endcsname\relax
\typeout{** WARNING: IEEEtran.bst: No hyphenation pattern has been}%
\typeout{** loaded for the language `#1'. Using the pattern for}%
\typeout{** the default language instead.}%
\else
\language=\csname l@#1\endcsname
\fi
#2}}
\providecommand{\BIBdecl}{\relax}
\BIBdecl

\bibitem{elias2021parallel}
I.~Elias, H.~Zen, J.~Shen, Y.~Zhang, Y.~Jia, R.~Skerry-Ryan, and Y.~Wu,
  ``Parallel tacotron 2: A non-autoregressive neural tts model with
  differentiable duration modeling,'' in \emph{Proc. Interspeech 2021}, 2021,
  pp. 141--145.

\bibitem{ren2020fastspeech}
Y.~Ren, C.~Hu, X.~Tan, T.~Qin, S.~Zhao, Z.~Zhao, and T.-Y. Liu, ``Fastspeech 2:
  Fast and high-quality end-to-end text to speech,'' in \emph{International
  Conference on Learning Representations}, 2021.

\bibitem{oord2016wavenet}
A.~v.~d. Oord, S.~Dieleman, H.~Zen, K.~Simonyan, O.~Vinyals, A.~Graves,
  N.~Kalchbrenner, A.~Senior, and K.~Kavukcuoglu, ``Wavenet: A generative model
  for raw audio,'' \emph{arXiv preprint arXiv:1609.03499}, 2016.

\bibitem{qian2020unsupervised}
K.~Qian, Y.~Zhang, S.~Chang, M.~Hasegawa-Johnson, and D.~Cox, ``Unsupervised
  speech decomposition via triple information bottleneck,'' in
  \emph{International Conference on Machine Learning}.\hskip 1em plus 0.5em
  minus 0.4em\relax PMLR, 2020, pp. 7836--7846.

\bibitem{wang2017tacotron}
Y.~Wang, R.~Skerry-Ryan, D.~Stanton, Y.~Wu, R.~J. Weiss, N.~Jaitly, Z.~Yang,
  Y.~Xiao, Z.~Chen, S.~Bengio, Q.~Le, Y.~Agiomyrgiannakis, R.~Clark, and R.~A.
  Saurous, ``Tacotron: Towards end-to-end speech synthesis,'' in \emph{Proc.
  Interspeech 2017}, 2017, pp. 4006--4010.

\bibitem{shen2018natural}
J.~Shen, R.~Pang, R.~J. Weiss, M.~Schuster, N.~Jaitly, Z.~Yang, Z.~Chen,
  Y.~Zhang, Y.~Wang, R.~Skerrv-Ryan \emph{et~al.}, ``Natural tts synthesis by
  conditioning wavenet on mel spectrogram predictions,'' in \emph{2018 IEEE
  International Conference on Acoustics, Speech and Signal Processing
  (ICASSP)}.\hskip 1em plus 0.5em minus 0.4em\relax IEEE, 2018, pp. 4779--4783.

\bibitem{ze2013statistical}
H.~Ze, A.~Senior, and M.~Schuster, ``Statistical parametric speech synthesis
  using deep neural networks,'' in \emph{2013 ieee international conference on
  acoustics, speech and signal processing}.\hskip 1em plus 0.5em minus
  0.4em\relax IEEE, 2013, pp. 7962--7966.

\bibitem{tokuda2013speech}
K.~Tokuda, Y.~Nankaku, T.~Toda, H.~Zen, J.~Yamagishi, and K.~Oura, ``Speech
  synthesis based on hidden markov models,'' \emph{Proceedings of the IEEE},
  vol. 101, no.~5, pp. 1234--1252, 2013.

\bibitem{kalchbrenner2018efficient}
N.~Kalchbrenner, E.~Elsen, K.~Simonyan, S.~Noury, N.~Casagrande, E.~Lockhart,
  F.~Stimberg, A.~Oord, S.~Dieleman, and K.~Kavukcuoglu, ``Efficient neural
  audio synthesis,'' in \emph{International Conference on Machine
  Learning}.\hskip 1em plus 0.5em minus 0.4em\relax PMLR, 2018, pp. 2410--2419.

\bibitem{morise2016world}
M.~Morise, F.~Yokomori, and K.~Ozawa, ``World: a vocoder-based high-quality
  speech synthesis system for real-time applications,'' \emph{IEICE
  TRANSACTIONS on Information and Systems}, vol.~99, no.~7, pp. 1877--1884,
  2016.

\bibitem{kawahara2001aperiodicity}
H.~Kawahara, J.~Estill, and O.~Fujimura, ``Aperiodicity extraction and control
  using mixed mode excitation and group delay manipulation for a high quality
  speech analysis, modification and synthesis system {STRAIGHT},'' in
  \emph{Second International Workshop on Models and Analysis of Vocal Emissions
  for Biomedical Applications}, 2001.

\bibitem{wu2016merlin}
Z.~Wu, O.~Watts, and S.~King, ``Merlin: An open source neural network speech
  synthesis system.'' in \emph{SSW}, 2016, pp. 202--207.

\bibitem{griffin1984signal}
D.~Griffin and J.~Lim, ``Signal estimation from modified short-time fourier
  transform,'' \emph{IEEE Transactions on Acoustics, Speech, and Signal
  Processing}, vol.~32, no.~2, pp. 236--243, 1984.

\bibitem{chou2018multi}
J.~chieh Chou, C.~chieh Yeh, H.~yi~Lee, and L.~shan Lee, ``Multi-target voice
  conversion without parallel data by adversarially learning disentangled audio
  representations,'' in \emph{Proc. Interspeech 2018}, 2018, pp. 501--505.

\bibitem{yeh2018rhythm}
C.-c. Yeh, P.-c. Hsu, J.-c. Chou, H.-y. Lee, and L.-s. Lee, ``Rhythm-flexible
  voice conversion without parallel data using cycle-gan over phoneme
  posteriorgram sequences,'' in \emph{2018 IEEE Spoken Language Technology
  Workshop (SLT)}.\hskip 1em plus 0.5em minus 0.4em\relax IEEE, 2018, pp.
  274--281.

\bibitem{liu2019unsupervised}
A.~T. Liu, P.~chun Hsu, and H.-Y. Lee, ``Unsupervised end-to-end learning of
  discrete linguistic units for voice conversion,'' in \emph{Proc. Interspeech
  2019}, 2019, pp. 1108--1112.

\bibitem{chien2021investigating}
C.-M. Chien, J.-H. Lin, C.-y. Huang, P.-c. Hsu, and H.-y. Lee, ``Investigating
  on incorporating pretrained and learnable speaker representations for
  multi-speaker multi-style text-to-speech,'' in \emph{ICASSP 2021-2021 IEEE
  International Conference on Acoustics, Speech and Signal Processing
  (ICASSP)}.\hskip 1em plus 0.5em minus 0.4em\relax IEEE, 2021, pp. 8588--8592.

\bibitem{qian2019autovc}
K.~Qian, Y.~Zhang, S.~Chang, X.~Yang, and M.~Hasegawa-Johnson, ``Autovc:
  Zero-shot voice style transfer with only autoencoder loss,'' in
  \emph{International Conference on Machine Learning}.\hskip 1em plus 0.5em
  minus 0.4em\relax PMLR, 2019, pp. 5210--5219.

\bibitem{ling2018waveform}
Z.-H. Ling, Y.~Ai, Y.~Gu, and L.-R. Dai, ``Waveform modeling and generation
  using hierarchical recurrent neural networks for speech bandwidth
  extension,'' \emph{IEEE/ACM Transactions on Audio, Speech, and Language
  Processing}, vol.~26, no.~5, pp. 883--894, 2018.

\bibitem{kleijn2021generative}
W.~B. Kleijn, A.~Storus, M.~Chinen, T.~Denton, F.~S. Lim, A.~Luebs,
  J.~Skoglund, and H.~Yeh, ``Generative speech coding with predictive variance
  regularization,'' in \emph{ICASSP 2021-2021 IEEE International Conference on
  Acoustics, Speech and Signal Processing (ICASSP)}.\hskip 1em plus 0.5em minus
  0.4em\relax IEEE, 2021, pp. 6478--6482.

\bibitem{polyak2021speech}
A.~Polyak, Y.~Adi, J.~Copet, E.~Kharitonov, K.~Lakhotia, W.-N. Hsu, A.~Mohamed,
  and E.~Dupoux, ``Speech resynthesis from discrete disentangled
  self-supervised representations,'' in \emph{Proc. Interspeech 2021}, 2021,
  pp. 3615--3619.

\bibitem{lorenzo2018towards}
J.~Lorenzo-Trueba, T.~Drugman, J.~Latorre, T.~Merritt, B.~Putrycz,
  R.~Barra-Chicote, A.~Moinet, and V.~Aggarwal, ``Towards achieving robust
  universal neural vocoding,'' in \emph{Proc. Interspeech 2019}, 2019, pp.
  181--185.

\bibitem{hsu2019towards}
P.-c. Hsu, C.-h. Wang, A.~T. Liu, and H.-y. Lee, ``Towards robust neural
  vocoding for speech generation: A survey,'' \emph{arXiv preprint
  arXiv:1912.02461}, 2019.

\bibitem{jiao2021universal}
Y.~Jiao, A.~Gabry{\'s}, G.~Tinchev, B.~Putrycz, D.~Korzekwa, and V.~Klimkov,
  ``Universal neural vocoding with parallel wavenet,'' in \emph{ICASSP
  2021-2021 IEEE International Conference on Acoustics, Speech and Signal
  Processing (ICASSP)}.\hskip 1em plus 0.5em minus 0.4em\relax IEEE, 2021, pp.
  6044--6048.

\bibitem{jin2018fftnet}
Z.~Jin, A.~Finkelstein, G.~J. Mysore, and J.~Lu, ``Fftnet: A real-time
  speaker-dependent neural vocoder,'' in \emph{2018 IEEE International
  Conference on Acoustics, Speech and Signal Processing (ICASSP)}.\hskip 1em
  plus 0.5em minus 0.4em\relax IEEE, 2018, pp. 2251--2255.

\bibitem{kanagawa2020lightweight}
H.~Kanagawa and Y.~Ijima, ``Lightweight lpcnet-based neural vocoder with tensor
  decomposition.'' in \emph{INTERSPEECH}, 2020, pp. 205--209.

\bibitem{valin2019lpcnet}
J.-M. Valin and J.~Skoglund, ``Lpcnet: Improving neural speech synthesis
  through linear prediction,'' in \emph{ICASSP 2019-2019 IEEE International
  Conference on Acoustics, Speech and Signal Processing (ICASSP)}.\hskip 1em
  plus 0.5em minus 0.4em\relax IEEE, 2019, pp. 5891--5895.

\bibitem{okamoto2017subband}
T.~Okamoto, K.~Tachibana, T.~Toda, Y.~Shiga, and H.~Kawai, ``Subband wavenet
  with overlapped single-sideband filterbanks,'' in \emph{2017 IEEE Automatic
  Speech Recognition and Understanding Workshop (ASRU)}.\hskip 1em plus 0.5em
  minus 0.4em\relax IEEE, 2017, pp. 698--704.

\bibitem{okamoto2018improving}
T.~Okamoto, T.~Toda, Y.~Shiga, and H.~Kawai, ``Improving fftnet vocoder with
  noise shaping and subband approaches,'' in \emph{2018 IEEE Spoken Language
  Technology Workshop (SLT)}.\hskip 1em plus 0.5em minus 0.4em\relax IEEE,
  2018, pp. 304--311.

\bibitem{yu2019durian}
C.~Yu, H.~Lu, N.~Hu, M.~Yu, C.~Weng, K.~Xu, P.~Liu, D.~Tuo, S.~Kang, G.~Lei,
  D.~Su, and D.~Yu, ``Durian: Duration informed attention network for speech
  synthesis,'' in \emph{Proc. Interspeech 2020}, 2020, pp. 2027--2031.

\bibitem{tian2020featherwave}
Q.~Tian, Z.~Zhang, H.~Lu, L.-H. Chen, and S.~Liu, ``Featherwave: An efficient
  high-fidelity neural vocoder with multi-band linear prediction,'' in
  \emph{Proc. Interspeech 2020}, 2020, pp. 195--199.

\bibitem{cui2020efficient}
Y.~Cui, X.~Wang, L.~He, and F.~K. Soong, ``An efficient subband linear
  prediction for lpcnet-based neural synthesis,'' in \emph{Proc. Interspeech
  2020}, 2020, pp. 3555--3559.

\bibitem{vipperla2020bunched}
R.~Vipperla, S.~Park, K.~Choo, S.~Ishtiaq, K.~Min, S.~Bhattacharya,
  A.~Mehrotra, A.~G.~C. Ramos, and N.~D. Lane, ``Bunched lpcnet: Vocoder for
  low-cost neural text-to-speech systems,'' in \emph{Proc. Interspeech 2020},
  2020, pp. 3565--3569.

\bibitem{oord2018parallel}
A.~Oord, Y.~Li, I.~Babuschkin, K.~Simonyan, O.~Vinyals, K.~Kavukcuoglu,
  G.~Driessche, E.~Lockhart, L.~Cobo, F.~Stimberg \emph{et~al.}, ``Parallel
  wavenet: Fast high-fidelity speech synthesis,'' in \emph{International
  conference on machine learning}.\hskip 1em plus 0.5em minus 0.4em\relax PMLR,
  2018, pp. 3918--3926.

\bibitem{yamamoto2020parallel}
R.~Yamamoto, E.~Song, and J.-M. Kim, ``Parallel wavegan: A fast waveform
  generation model based on generative adversarial networks with
  multi-resolution spectrogram,'' in \emph{ICASSP 2020-2020 IEEE International
  Conference on Acoustics, Speech and Signal Processing (ICASSP)}.\hskip 1em
  plus 0.5em minus 0.4em\relax IEEE, 2020, pp. 6199--6203.

\bibitem{ping2018clarinet}
W.~Ping, K.~Peng, and J.~Chen, ``Clarinet: Parallel wave generation in
  end-to-end text-to-speech,'' in \emph{International Conference on Learning
  Representations}, 2019.

\bibitem{prenger2019waveglow}
R.~Prenger, R.~Valle, and B.~Catanzaro, ``Waveglow: A flow-based generative
  network for speech synthesis,'' in \emph{ICASSP 2019-2019 IEEE International
  Conference on Acoustics, Speech and Signal Processing (ICASSP)}.\hskip 1em
  plus 0.5em minus 0.4em\relax IEEE, 2019, pp. 3617--3621.

\bibitem{kim2018flowavenet}
S.~Kim, S.-G. Lee, J.~Song, J.~Kim, and S.~Yoon, ``Flowavenet: A generative
  flow for raw audio,'' in \emph{International Conference on Machine
  Learning}.\hskip 1em plus 0.5em minus 0.4em\relax PMLR, 2019, pp. 3370--3378.

\bibitem{ping2020waveflow}
W.~Ping, K.~Peng, K.~Zhao, and Z.~Song, ``Waveflow: A compact flow-based model
  for raw audio,'' in \emph{International Conference on Machine
  Learning}.\hskip 1em plus 0.5em minus 0.4em\relax PMLR, 2020, pp. 7706--7716.

\bibitem{kumar2019melgan}
K.~Kumar, R.~Kumar, T.~De~Boissiere, L.~Gestin, W.~Z. Teoh, J.~Sotelo,
  A.~de~Br{\'e}bisson, Y.~Bengio, and A.~C. Courville, ``Melgan: Generative
  adversarial networks for conditional waveform synthesis,'' \emph{Advances in
  neural information processing systems}, vol.~32, 2019.

\bibitem{kong2020hifi}
J.~Kong, J.~Kim, and J.~Bae, ``Hifi-gan: Generative adversarial networks for
  efficient and high fidelity speech synthesis,'' \emph{Advances in Neural
  Information Processing Systems}, vol.~33, pp. 17\,022--17\,033, 2020.

\bibitem{chen2020wavegrad}
N.~Chen, Y.~Zhang, H.~Zen, R.~J. Weiss, M.~Norouzi, and W.~Chan, ``Wavegrad:
  Estimating gradients for waveform generation,'' in \emph{International
  Conference on Learning Representations}, 2021.

\bibitem{kong2020diffwave}
Z.~Kong, W.~Ping, J.~Huang, K.~Zhao, and B.~Catanzaro, ``Diffwave: A versatile
  diffusion model for audio synthesis,'' in \emph{International Conference on
  Learning Representations}, 2021.

\bibitem{hsu2020wg}
P.-c. Hsu and H.-y. Lee, ``Wg-wavenet: Real-time high-fidelity speech synthesis
  without gpu,'' \emph{Proc. Interspeech 2020}, pp. 210--214, 2020.

\bibitem{song2021improved}
E.~Song, R.~Yamamoto, M.-J. Hwang, J.-S. Kim, O.~Kwon, and J.-M. Kim,
  ``Improved parallel wavegan vocoder with perceptually weighted spectrogram
  loss,'' in \emph{2021 IEEE Spoken Language Technology Workshop (SLT)}.\hskip
  1em plus 0.5em minus 0.4em\relax IEEE, 2021, pp. 470--476.

\bibitem{gritsenko2020spectral}
A.~Gritsenko, T.~Salimans, R.~van~den Berg, J.~Snoek, and N.~Kalchbrenner, ``A
  spectral energy distance for parallel speech synthesis,'' \emph{Advances in
  Neural Information Processing Systems}, vol.~33, pp. 13\,062--13\,072, 2020.

\bibitem{vaswani2017attention}
A.~Vaswani, N.~Shazeer, N.~Parmar, J.~Uszkoreit, L.~Jones, A.~N. Gomez,
  {\L}.~Kaiser, and I.~Polosukhin, ``Attention is all you need,''
  \emph{Advances in neural information processing systems}, vol.~30, 2017.

\bibitem{brown2020language}
T.~Brown, B.~Mann, N.~Ryder, M.~Subbiah, J.~D. Kaplan, P.~Dhariwal,
  A.~Neelakantan, P.~Shyam, G.~Sastry, A.~Askell \emph{et~al.}, ``Language
  models are few-shot learners,'' \emph{Advances in neural information
  processing systems}, vol.~33, pp. 1877--1901, 2020.

\bibitem{lewis2019bart}
M.~Lewis, Y.~Liu, N.~Goyal, M.~Ghazvininejad, A.~Mohamed, O.~Levy, V.~Stoyanov,
  and L.~Zettlemoyer, ``Bart: Denoising sequence-to-sequence pre-training for
  natural language generation, translation, and comprehension,'' in
  \emph{Proceedings of the 58th Annual Meeting of the Association for
  Computational Linguistics}, 2020, pp. 7871--7880.

\bibitem{van2016pixel}
A.~Van~Oord, N.~Kalchbrenner, and K.~Kavukcuoglu, ``Pixel recurrent neural
  networks,'' in \emph{International conference on machine learning}.\hskip 1em
  plus 0.5em minus 0.4em\relax PMLR, 2016, pp. 1747--1756.

\bibitem{van2016conditional}
A.~Van~den Oord, N.~Kalchbrenner, L.~Espeholt, O.~Vinyals, A.~Graves
  \emph{et~al.}, ``Conditional image generation with pixelcnn decoders,''
  \emph{Advances in neural information processing systems}, vol.~29, 2016.

\bibitem{yu2022scaling}
J.~Yu, Y.~Xu, J.~Y. Koh, T.~Luong, G.~Baid, Z.~Wang, V.~Vasudevan, A.~Ku,
  Y.~Yang, B.~K. Ayan, B.~Hutchinson, W.~Han, Z.~Parekh, X.~Li, H.~Zhang,
  J.~Baldridge, and Y.~Wu, ``Scaling autoregressive models for content-rich
  text-to-image generation,'' \emph{Transactions on Machine Learning Research},
  2022.

\bibitem{nash2020polygen}
C.~Nash, Y.~Ganin, S.~A. Eslami, and P.~Battaglia, ``Polygen: An autoregressive
  generative model of 3d meshes,'' in \emph{International conference on machine
  learning}.\hskip 1em plus 0.5em minus 0.4em\relax PMLR, 2020, pp. 7220--7229.

\bibitem{yang2021multi}
G.~Yang, S.~Yang, K.~Liu, P.~Fang, W.~Chen, and L.~Xie, ``Multi-band melgan:
  Faster waveform generation for high-quality text-to-speech,'' in \emph{2021
  IEEE Spoken Language Technology Workshop (SLT)}.\hskip 1em plus 0.5em minus
  0.4em\relax IEEE, 2021, pp. 492--498.

\bibitem{ming2002robust}
J.~Ming, P.~Jancovic, and F.~J. Smith, ``Robust speech recognition using
  probabilistic union models,'' \emph{IEEE Transactions on Speech and Audio
  Processing}, vol.~10, no.~6, pp. 403--414, 2002.

\bibitem{mcauley2005subband}
J.~McAuley, J.~Ming, D.~Stewart, and P.~Hanna, ``Subband correlation and robust
  speech recognition,'' \emph{IEEE Transactions on Speech and Audio
  Processing}, vol.~13, no.~5, pp. 956--964, 2005.

\bibitem{piao2007image}
Y.~Piao, H.~Park \emph{et~al.}, ``Image resolution enhancement using
  inter-subband correlation in wavelet domain,'' in \emph{2007 IEEE
  International Conference on Image Processing}, vol.~1.\hskip 1em plus 0.5em
  minus 0.4em\relax IEEE, 2007, pp. I--445.

\bibitem{bhuiyan2014subband}
M.~I.~H. Bhuiyan and A.~B. Das, ``A subband correlation-based method for the
  automatic detection of epilepsy and seizure in the dual tree complex wavelet
  transform domain,'' in \emph{2014 IEEE Conference on Biomedical Engineering
  and Sciences (IECBES)}.\hskip 1em plus 0.5em minus 0.4em\relax IEEE, 2014,
  pp. 811--816.

\bibitem{oord2016conditional}
A.~Van~den Oord, N.~Kalchbrenner, L.~Espeholt, O.~Vinyals, A.~Graves
  \emph{et~al.}, ``Conditional image generation with pixelcnn decoders,''
  \emph{Advances in neural information processing systems}, vol.~29, 2016.

\bibitem{kingma2016improving}
D.~P. Kingma, T.~Salimans, R.~Jozefowicz, X.~Chen, I.~Sutskever, and
  M.~Welling, ``Improved variational inference with inverse autoregressive
  flow,'' in \emph{Advances in Neural Information Processing Systems}, vol.~29,
  2016.

\bibitem{kingma2018glow}
D.~P. Kingma and P.~Dhariwal, ``Glow: Generative flow with invertible 1x1
  convolutions,'' \emph{Advances in neural information processing systems},
  vol.~31, 2018.

\bibitem{gu2018non}
J.~Gu, J.~Bradbury, C.~Xiong, V.~O. Li, and R.~Socher, ``Non-autoregressive
  neural machine translation,'' in \emph{International Conference on Learning
  Representations}, 2018.

\bibitem{lee2020deterministic}
J.~Lee, E.~Mansimov, and K.~Cho, ``Deterministic non-autoregressive neural
  sequence modeling by iterative refinement,'' in \emph{2018 Conference on
  Empirical Methods in Natural Language Processing, EMNLP 2018}.\hskip 1em plus
  0.5em minus 0.4em\relax Association for Computational Linguistics, 2018, pp.
  1173--1182.

\bibitem{su2021non}
Y.~Su, D.~Cai, Y.~Wang, D.~Vandyke, S.~Baker, P.~Li, and N.~Collier,
  ``Non-autoregressive text generation with pre-trained language models,'' in
  \emph{Proceedings of the 16th Conference of the European Chapter of the
  Association for Computational Linguistics: Main Volume}, 2021, pp. 234--243.

\bibitem{xiao2023survey}
Y.~Xiao, L.~Wu, J.~Guo, J.~Li, M.~Zhang, T.~Qin, and T.-y. Liu, ``A survey on
  non-autoregressive generation for neural machine translation and beyond,''
  \emph{IEEE Transactions on Pattern Analysis and Machine Intelligence}, 2023.

\bibitem{ren2020study}
Y.~Ren, J.~Liu, X.~Tan, Z.~Zhao, S.~Zhao, and T.-Y. Liu, ``A study of
  non-autoregressive model for sequence generation,'' in \emph{Proceedings of
  the 58th Annual Meeting of the Association for Computational Linguistics},
  2020, pp. 149--159.

\bibitem{wu2018beyond}
L.~Wu, X.~Tan, D.~He, F.~Tian, T.~Qin, J.~Lai, and T.-Y. Liu, ``Beyond error
  propagation in neural machine translation: Characteristics of language also
  matter,'' in \emph{Proceedings of the 2018 Conference on Empirical Methods in
  Natural Language Processing}, 2018, pp. 3602--3611.

\bibitem{morrison2022chunked}
M.~Morrison, R.~Kumar, K.~Kumar, P.~Seetharaman, A.~Courville, and Y.~Bengio,
  ``Chunked autoregressive {GAN} for conditional waveform synthesis,'' in
  \emph{International Conference on Learning Representations}, 2022.

\bibitem{nguyen1994near}
T.~Nguyen, ``Near-perfect-reconstruction pseudo-qmf banks,'' \emph{IEEE
  Transactions on Signal Processing}, vol.~42, no.~1, pp. 65--76, 1994.

\bibitem{chorowski2019unsupervised}
J.~Chorowski, R.~J. Weiss, S.~Bengio, and A.~van~den Oord, ``Unsupervised
  speech representation learning using wavenet autoencoders,'' \emph{IEEE/ACM
  transactions on audio, speech, and language processing}, vol.~27, no.~12, pp.
  2041--2053, 2019.

\bibitem{rethage2018wavenet}
D.~Rethage, J.~Pons, and X.~Serra, ``A wavenet for speech denoising,'' in
  \emph{2018 IEEE International Conference on Acoustics, Speech and Signal
  Processing (ICASSP)}.\hskip 1em plus 0.5em minus 0.4em\relax IEEE, 2018, pp.
  5069--5073.

\bibitem{misra2019mish}
D.~Misra, ``Mish: A self regularized non-monotonic neural activation
  function,'' \emph{arXiv preprint arXiv:1908.08681}, 2019.

\bibitem{xie2008dynamic}
C.~S. Xie, ``Dynamic vs static autoregressive models for forecasting time
  series,'' \emph{Available at SSRN 1268910}, 2008.

\bibitem{wadhvani2017review}
R.~Wadhvani \emph{et~al.}, ``Review on various models for time series
  forecasting,'' in \emph{2017 International Conference on Inventive Computing
  and Informatics (ICICI)}.\hskip 1em plus 0.5em minus 0.4em\relax IEEE, 2017,
  pp. 405--410.

\bibitem{takaki2019stft}
S.~Takaki, T.~Nakashika, X.~Wang, and J.~Yamagishi, ``Stft spectral loss for
  training a neural speech waveform model,'' in \emph{ICASSP 2019-2019 IEEE
  International Conference on Acoustics, Speech and Signal Processing
  (ICASSP)}.\hskip 1em plus 0.5em minus 0.4em\relax IEEE, 2019, pp. 7065--7069.

\bibitem{arik2018fast}
S.~{\"O}. Ar{\i}k, H.~Jun, and G.~Diamos, ``Fast spectrogram inversion using
  multi-head convolutional neural networks,'' \emph{IEEE Signal Processing
  Letters}, vol.~26, no.~1, pp. 94--98, 2018.

\bibitem{tian2020tfgan}
Q.~Tian, Y.~Chen, Z.~Zhang, H.~Lu, L.~Chen, L.~Xie, and S.~Liu, ``Tfgan: Time
  and frequency domain based generative adversarial network for high-fidelity
  speech synthesis,'' \emph{arXiv preprint arXiv:2011.12206}, 2020.

\bibitem{ljspeech17}
K.~Ito and L.~Johnson, ``The lj speech dataset,''
  \url{https://keithito.com/LJ-Speech-Dataset/}, 2017.

\bibitem{yamagishi2017cstr}
J.~Yamagishi, C.~Veaux, K.~MacDonald \emph{et~al.}, ``Cstr vctk corpus: English
  multi-speaker corpus for cstr voice cloning toolkit,'' 2017.

\bibitem{kominek2004cmu}
J.~Kominek and A.~W. Black, ``The cmu arctic speech databases,'' in \emph{Fifth
  ISCA workshop on speech synthesis}, 2004.

\bibitem{kingma2014adam}
D.~P. Kingma and J.~Ba, ``Adam: A method for stochastic optimization,'' in
  \emph{International Conference on Learning Representations}, 2015.

\bibitem{yang2020vocgan}
J.~Yang, J.~Lee, Y.~Kim, H.-Y. Cho, and I.~Kim, ``Vocgan: A high-fidelity
  real-time vocoder with a hierarchically-nested adversarial network,'' in
  \emph{Proc. Interspeech 2020}, 2020, pp. 200--204.

\bibitem{Ranger}
L.~Wright, ``Ranger - a synergistic optimizer.''
  \url{https://github.com/lessw2020/Ranger-Deep-Learning-Optimizer}, 2019.

\bibitem{zhai2020squeezewave}
B.~Zhai, T.~Gao, F.~Xue, D.~Rothchild, B.~Wu, J.~E. Gonzalez, and K.~Keutzer,
  ``Squeezewave: Extremely lightweight vocoders for on-device speech
  synthesis,'' \emph{arXiv preprint arXiv:2001.05685}, 2020.

\bibitem{kubichek1993mel}
R.~Kubichek, ``Mel-cepstral distance measure for objective speech quality
  assessment,'' in \emph{Proceedings of IEEE Pacific Rim Conference on
  Communications Computers and Signal Processing}, vol.~1.\hskip 1em plus 0.5em
  minus 0.4em\relax IEEE, 1993, pp. 125--128.

\bibitem{mauch2014pyin}
M.~Mauch and S.~Dixon, ``pyin: A fundamental frequency estimator using
  probabilistic threshold distributions,'' in \emph{2014 ieee international
  conference on acoustics, speech and signal processing (icassp)}.\hskip 1em
  plus 0.5em minus 0.4em\relax IEEE, 2014, pp. 659--663.

\bibitem{recommendation2001method}
I.~Recommendation, ``Method for the subjective assessment of intermediate sound
  quality (mushra),'' \emph{ITU, BS}, pp. 1543--1, 2001.

\bibitem{merritt2018comprehensive}
T.~Merritt, B.~Putrycz, A.~Nadolski, T.~Ye, D.~Korzekwa, W.~Dolecki,
  T.~Drugman, V.~Klimkov, A.~Moinet, A.~Breen \emph{et~al.}, ``Comprehensive
  evaluation of statistical speech waveform synthesis,'' in \emph{2018 IEEE
  Spoken Language Technology Workshop (SLT)}.\hskip 1em plus 0.5em minus
  0.4em\relax IEEE, 2018, pp. 325--331.

\bibitem{cotescu2019voice}
M.~Cotescu, T.~Drugman, G.~Huybrechts, J.~Lorenzo-Trueba, and A.~Moinet,
  ``Voice conversion for whispered speech synthesis,'' \emph{IEEE Signal
  Processing Letters}, vol.~27, pp. 186--190, 2019.

\bibitem{latorre2019effect}
J.~Latorre, J.~Lachowicz, J.~Lorenzo-Trueba, T.~Merritt, T.~Drugman,
  S.~Ronanki, and V.~Klimkov, ``Effect of data reduction on
  sequence-to-sequence neural tts,'' in \emph{ICASSP 2019-2019 IEEE
  International Conference on Acoustics, Speech and Signal Processing
  (ICASSP)}.\hskip 1em plus 0.5em minus 0.4em\relax IEEE, 2019, pp. 7075--7079.

\bibitem{gabrys2022voice}
A.~Gabry{\'s}, G.~Huybrechts, M.~S. Ribeiro, C.-M. Chien, J.~Roth, G.~Comini,
  R.~Barra-Chicote, B.~Perz, and J.~Lorenzo-Trueba, ``Voice filter: Few-shot
  text-to-speech speaker adaptation using voice conversion as a post-processing
  module,'' in \emph{ICASSP 2022-2022 IEEE International Conference on
  Acoustics, Speech and Signal Processing (ICASSP)}.\hskip 1em plus 0.5em minus
  0.4em\relax IEEE, 2022, pp. 7902--7906.

\bibitem{deja22_interspeech}
K.~Deja, A.~Sanchez, J.~Roth, and M.~Cotescu, ``{Automatic Evaluation of
  Speaker Similarity},'' in \emph{Proc. Interspeech 2022}, 2022, pp.
  2348--2352.

\bibitem{goodfellow2014generative}
I.~Goodfellow, J.~Pouget-Abadie, M.~Mirza, B.~Xu, D.~Warde-Farley, S.~Ozair,
  A.~Courville, and Y.~Bengio, ``Generative adversarial nets,'' \emph{Advances
  in neural information processing systems}, vol.~27, 2014.

\bibitem{rothauser1969ieee}
E.~Rothauser, ``Ieee recommended practice for speech quality measurements,''
  \emph{IEEE Trans. on Audio and Electroacoustics}, vol.~17, pp. 225--246,
  1969.

\end{thebibliography}

\begin{IEEEbiography}[{\includegraphics[width=1in,height=1.25in,clip,keepaspectratio]{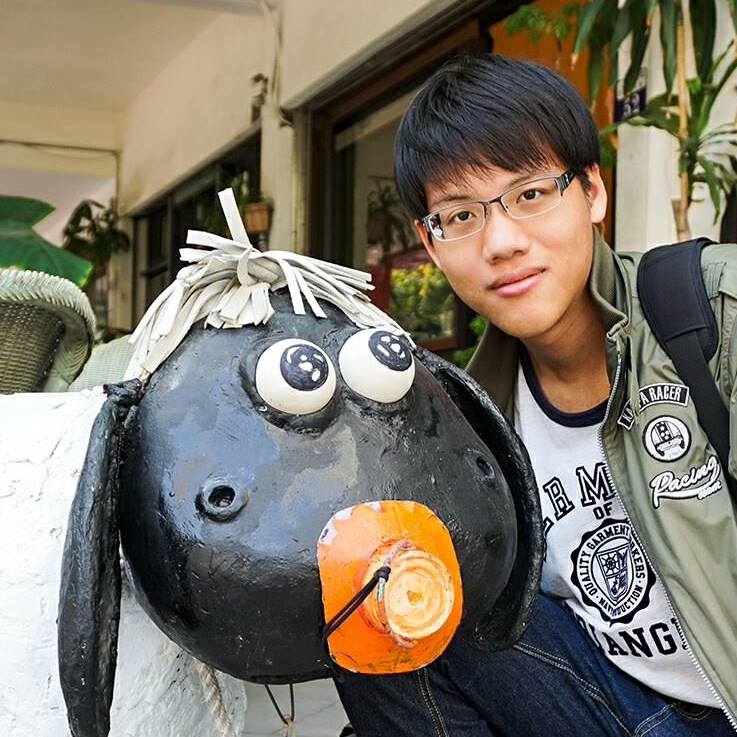}}]{Po-chun Hsu}
received the B.S. degree from National Taiwan University (NTU) in 2018 and is now a PhD student at the Graduate Institute of Communication Engineering (GICE) at NTU. His research focuses on speech synthesis, including text-to-speech (TTS), voice conversion (VC), and neural vocoder.
\end{IEEEbiography}

\begin{IEEEbiography}[{\includegraphics[width=1in,height=1.25in,clip,keepaspectratio]{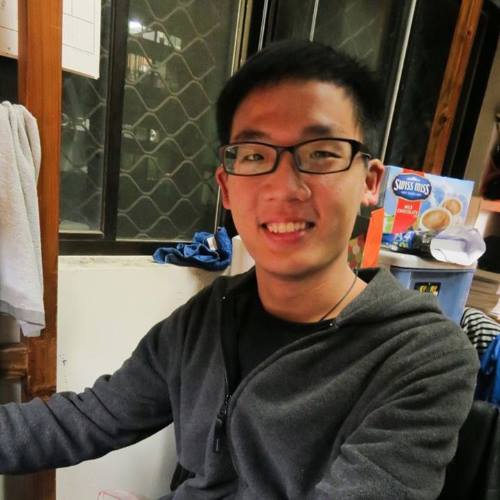}}]{Da-rong Liu}
Da-rong Liu, received the Bachelor degree from National Taiwan University (NTU) in 2016, and is now a PhD student at the Graduate Institute of Communication Engineering (GICE) at National Taiwan University. He mainly works on unsupervised learning, speech recognition and speech generation.
\end{IEEEbiography}

\begin{IEEEbiography}[{\includegraphics[width=1in,height=1.25in,clip,keepaspectratio]{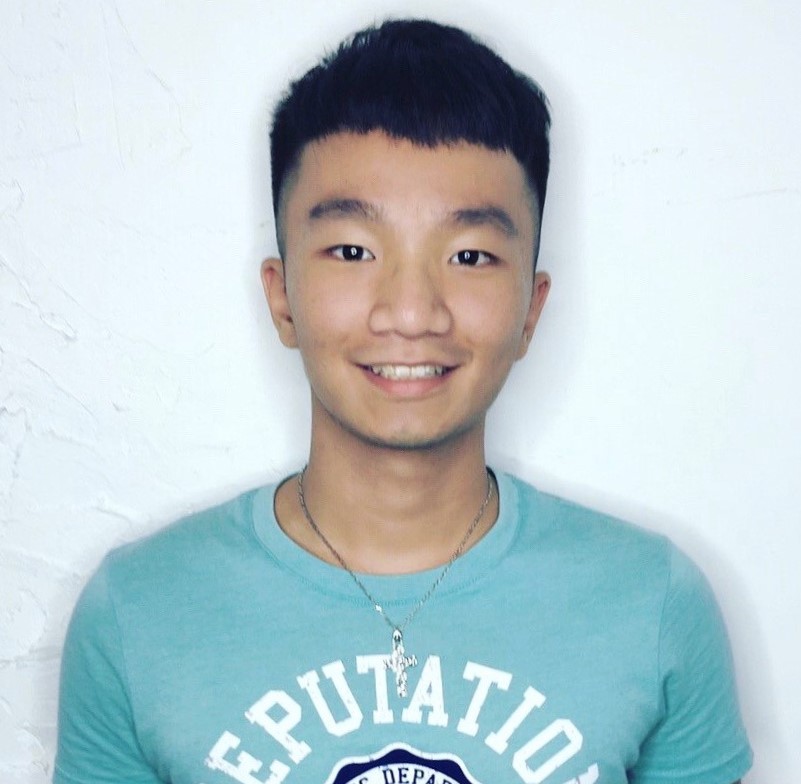}}]{Andy T. Liu}
received the bachelor’s degree in electrical engineering from National Taiwan University (NTU), Taipei, Taiwan, in 2018. He is currently working toward the PhD degree with the Graduate Institute of Communication Engineering, NTU, supervised by Professor Hung-yi Lee. His research interests include self-supervised learning, few-shot learning, and machine learning in the speech and NLP domain.
\end{IEEEbiography}

\begin{IEEEbiography}[{\includegraphics[width=1in,height=1.25in,clip,keepaspectratio]{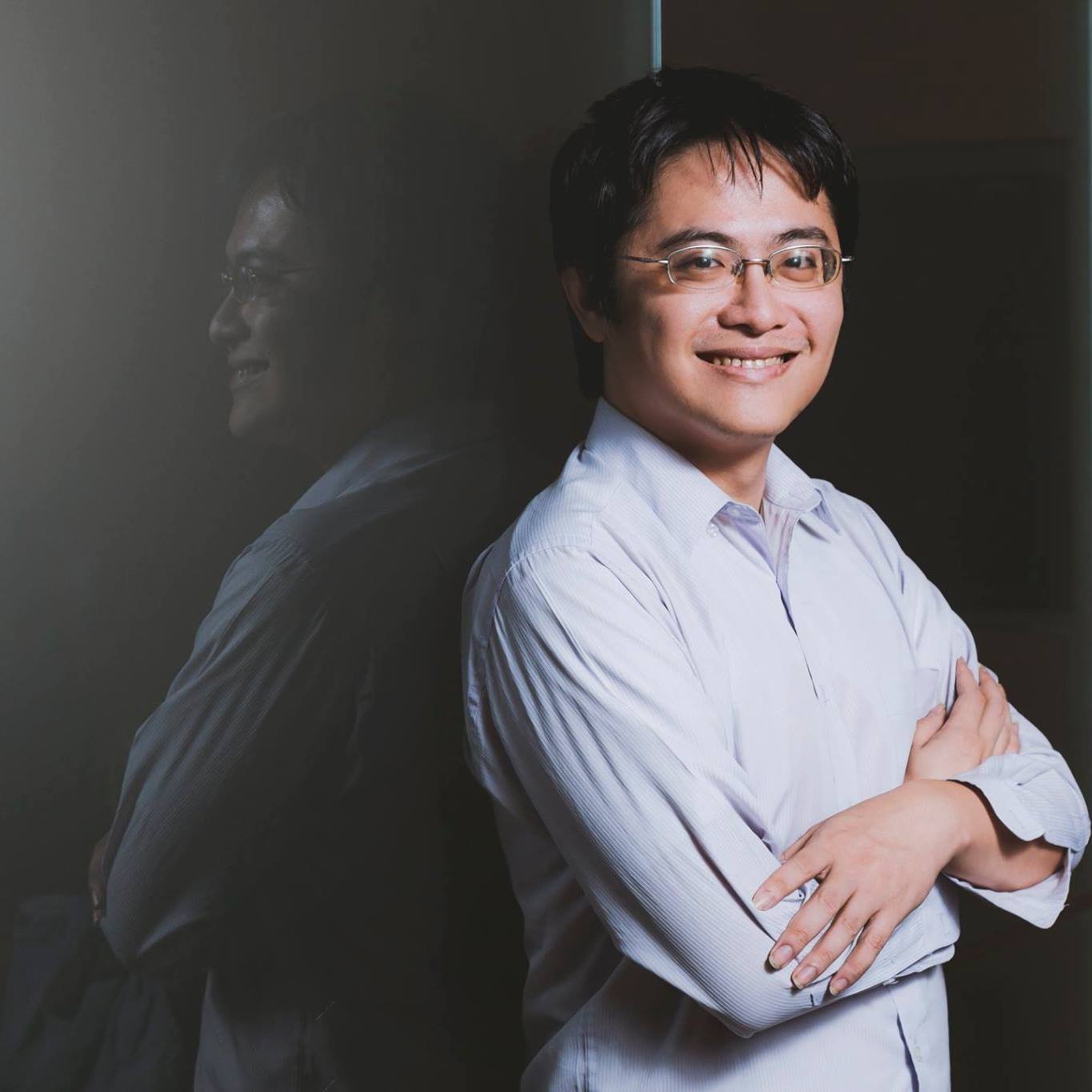}}]{Hung-yi Lee}
received the M.S. and Ph.D. degrees from National Taiwan University, Taipei, Taiwan, in 2010 and 2012, respectively. From September 2012 to August 2013, he was a Postdoctoral Fellow with Research Center for Information Technology Innovation, Academia Sinica. From September 2013 to July 2014, he was a Visiting Scientist with the Spoken Language Systems Group, MIT Computer Science and Artificial Intelligence Laboratory. He is currently an Associate Professor with the Department of Electrical Engineering, National Taiwan University, Taipei, Taiwan with a joint appointment to the Department of Computer Science and Information Engineering. His research focuses on spoken language understanding, speech recognition, and machine learning.
\end{IEEEbiography}

\end{document}